\documentclass[%
 reprint,
 amsmath,amssymb,
 aps,
prb,
]{revtex4-1}

\usepackage{graphicx}
\usepackage{dcolumn}
\usepackage{bm}
\usepackage{multirow}
\usepackage{xcolor}

\begin{document}


\title{Zirconia and hafnia polymorphs -- ground state structural properties from diffusion Monte Carlo}

\author{Hyeondeok Shin}%
\email{hshin@anl.gov}
\affiliation{Computational Sciences Division, Argonne National Laboratory, Argonne, IL 60439, United States}
\author{Anouar Benali}
\author{Ye Luo}
\affiliation{Argonne Leadership Computing Facility, Argonne National Laboratory, Lemont, IL 60439}
\affiliation{Computational Sciences Division, Argonne National Laboratory, Argonne, IL 60439, United States}

\author{Emily Crabb}
\affiliation{Department of Physics, Massachusetts Institute of Technology, Cambridge, MA 02139}
  \author{Alejandro Lopez-Bezanilla}
\affiliation{Materials Science Division, Argonne National Laboratory, Lemont, IL 60439}
\author{Laura E. Ratcliff}
\affiliation{Argonne Leadership Computing Facility, Argonne National Laboratory, Lemont, IL 60439
}
\affiliation{Department of Materials, Imperial College London, London SW7 2AZ, UK}
\author{Andrea M. Jokisaari}
\affiliation{Northwestern-Argonne Institute of Science and Engineering, Northwestern University, Evanston, IL 60208}
\author{Olle Heinonen}
\email{heinonen@anl.gov}
\affiliation{Materials Science Division, Argonne National Laboratory, Lemont, IL 60439}
\affiliation{Northwestern-Argonne Institute of Science and Engineering, Northwestern University, Evanston, IL 60208}
\email{heinonen@anl.gov}

\date{\today}

\begin{abstract}
Zirconia (zirconium dioxide) and hafnia (hafnium dioxide) are binary oxides used in a range of applications. Because zirconium and hafnium are chemically equivalent, they have three similar polymorphs, and it is important to understand  the properties and energetics of these polymorphs. However, while density functional theory calculations can get the correct energetic ordering, the energy differences between polymorphs depend very much on the specific density functional theory approach, as do other quantities such as lattice constants and bulk modulus. We have used highly accurate quantum Monte Carlo simulations to model the three zirconia and hafnia polymorphs. We compare our results for structural parameters, bulk modulus, and cohesive energy with results obtained from density functional theory calculations. We also discuss comparisons of our results with existing experimental data, in particular for structural parameters where extrapolation to zero temperature can be attempted. We hope our results of structural parameters as well as for cohesive energy and bulk modulus can serve as benchmarks for density-functional theory based calculations and as a guidance for future experiments.
\end{abstract}

\pacs{Valid PACS appear here}
\maketitle


\section{\label{sec:level1}Introduction}

Zirconium dioxide (zirconia), ZrO$_2$ is a simple oxide with a range of interesting properties that makes it useful for important current as well as potential future  applications. 
Zirconia has high mechanical strength and stability at elevated temperatures, high wear resistance, and is chemically inert. 
While originally used in refractory applications, it now also has application in wide range of areas, ranging from medical devices to cutting tools and solid electrolytes.\cite{manicone2007,bocanegra2002} 
Zirconia is also interesting -- and important -- for another reason. Zirconium alloys are used as cladding in nuclear fuel rods in nuclear power stations. During fuel burn-up with the fuel rods immersed in water, the zirconium will oxidize because of contact with the water, and hydrogen migrates into the zirconium metal alloy. 
The oxide is mostly protective in that it prevents water from being in direct contact with metallic zirconium, and further oxidation depends on diffusion of oxygen and hydrogen through the zirconia. However, the zirconia tends to crack along grain boundaries, at which oxidation continues unabated.\cite{motta2015}

Hafnium dioxide (hafnia), HfO$_2$, is also of great interest because of its unique electronic and structural properties.
Its wide band-gap, high thermal stability, and large dielectric constant 
make HfO$_2$ thin films important in applications such as 
optical coatings, and resistive random-access memory.\cite{saxena75,edlou93,lee00,kang00,kuhaili04,robertson06,khoshman08,kittl09,choi11,wong12}
Hafnia also has a high dielectric permittivity, good chemical compatibility with silicon, and a higher heat of formation than SiO$_2$. This makes hafnia ideal to replace SiO$_2$ in integrated electronic devices. Its high dielectric response enables reduction of the gate thickness of the gate dielectric layer in oxide-semiconductor field effect transistors while suppressing leakage currents through quantum mechanical tunneling through the dielectric layer.\cite{choi11} 
Because thin films of hafnia may exhibit a wide range of crystallographic phases and size-dependent phase transitions across several polymorphs, an accurate energetic analysis of the different crystalline forms with the best level of theory at hand is crucial for their theoretical characterization.

\begin{figure}[t]
 \includegraphics[width=3.0in]{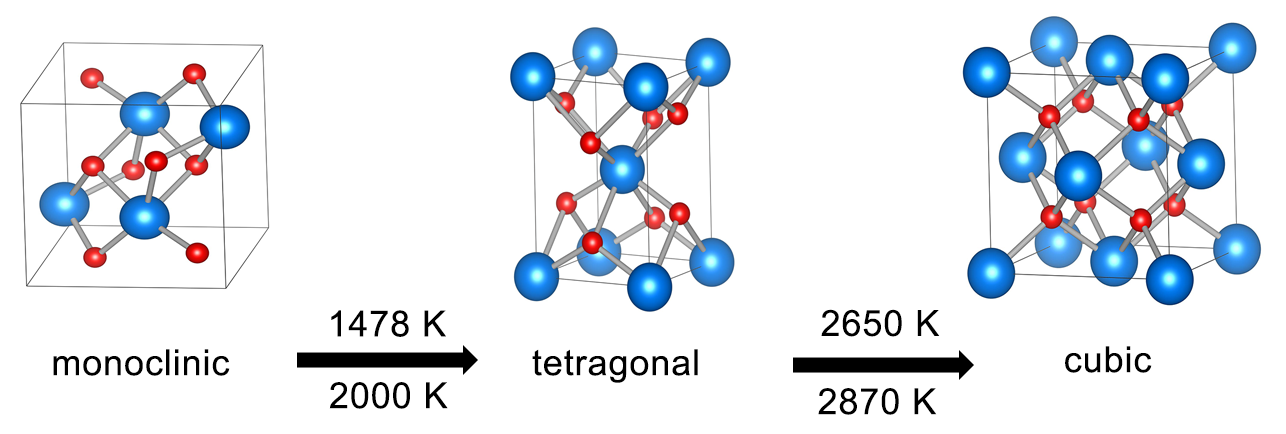}
 \caption{Structures of the three zirconia and hafnia polymorphs with oxygen in red and Zr or Hf in blue. The Zr (Hf) transition temperatures are indicated above (below) the arrows.}
 \label{fig:Phase_diag}
\end{figure}

Zirconium and hafnium are in the same column, column 4, of the periodic table and therefore have very similar chemical properties. They have similar ionic and atomic radii due to the lanthanide contraction;
the almost identical bonding nature and electronic properties of the ZrO$_2$ and HfO$_2$ molecules stem from the similarities of the atomic properties of Zr and Hf.\cite{chertihin95,brugh99,lesarri12}
Another consequence of the nearly indistinguishable chemical natures of the Hf and Zr atoms, is that bulk hafnia and zirconia have similar crystal phases and phase diagrams\cite{kisi98,massalski90,baun63,wolten63}. 
However, experimental and theoretical studies of the Si:HfO$_2$(ZrO$_2$) interface have shown that the Si:HfO$_2$ is more thermodynamically stable than the Si:ZrO$_2$ interface.\cite{gutowski02} 
This indicates that there are some differences in the {\em bulk} properties of ZrO$_2$ and HfO$_2$, in contrast to the very similar zirconia and hafnia  {\em molecular} systems. 
Zirconia and hafnia both have several polymorphs: at ambient pressure a low-temperature monoclinic (m-ZrO$_2$ and m-HfO$_2$) phase, a subsequent transformation to a tetragonal phase (t-ZrO$_2$ and t-HfO$_2$) at 1478~K for ZrO$_2$ and about 2000~K for HfO$_2$, and to a high-temperature cubic phase at (c-ZrO$_2$ and c-HfO$_2$) at 2650~K and 2870~K, respectively, as illustrated in Fig.~\ref{fig:Phase_diag}; the melting temperatures are about 2950~K and 3118~K, respectively.

For ZrO$_2$, the m-ZrO$_2$ to t-ZrO$_2$ transition is particularly important as this can bring about catastrophic fracture because of the accompanying shear strain of about 0.16 and 4\% volume change at the transition. Such strain can be accommodated by metals, but usually not by ceramics. However, it was realized that this transformation can be controlled and used as a source of transformation plasticity and transformation-toughening in engineered two-phase microstructures. This led to much expanded applications from its original limited use in refractory applications.
Because of the large volume changes not only in m-ZrO$_2$ to t-ZrO$_2$, but also in t-ZrO$_2$ to c-ZrO$_2$ (3\%), and their implications for applications it is of fundamental interest to understand the energetics of these phases. 

While there exist accurate experimental data for structural and electronic properties, such as lattice parameters, bulk modulus, and cohesive energy, for zirconia polymorphs obtained using X-ray or mass spectrometric measurements, there does not exist a similar large body of experimental data for hafnia.\cite{ackermann1975,aldebert85,howard88,stefanovich1994} 
Structural and electronic properties of the low-temperature m-HfO$_2$ phase have been determined at room temperature using X-ray diffraction.\cite{curtis54,adam59,ruh68,stacy72,ruh73,hann85,morales10} Lattice parameters have been reported for t-HfO$_2$ and c-HfO$_2$, but detailed structural and electronic characterizations are still lacking 
.\cite{curtis54,adam59,shanshoury70,wang92}
Because of difficulties in experimental measurements of the high-temperature HfO$_2$ polymorphs, it is necessary to use a theoretically approach to accurately establish structural and electronic properties of t-HfO$_2$ and c-HfO$_2$.

From a modeling perspective, the question is then how to accurately obtain the energetic ordering and stability of the zirconia polymorphs, as well as the basic structural and electronic properties of hafnia polymorphs. 
There have been a number of density functional theory-based studies of the energetics of the zirconia polymorphs\cite{dewhurst1998,gallino2011,jiang2010,dash2004,stefanovich1994,milman2009,jansen1991,kralik1998,finnis1998,terki2006,bredow2007,daramola2010,gallino2011,ricca2015,carbogno2014}; Jiang {\em et al.} also included many-body perturbation theory ($GW$-approximation) in their study of the electronic properties of zirconia\cite{jiang2010}. While DFT can get the energetic ordering right\cite{jansen1991,kralik1998,finnis1998}, the calculated energy differences between the polymorphs depend very much on the specific exchange-correlation functional that is used. 

Several works have attempted to determine the equation of state (EOS) of hafnia polymorphs using density functional theory (DFT) within the local density approximation (LDA) or generalized gradient approximation (GGA).\cite{lowther99,kang03,jaffe05} However, neither LDA nor GGA exchange-correlation (XC) functionals can simultaneously obtain good results for lattice parameter and bulk modulus; for example, in Ref.~\onlinecite{lowther99}, the LDA lattice parameters $a$, $b$, and $c$ for m-HfO$_2$ were computed to be 5.12, 5.17, and 5.29~$\text{\AA}$, respectively, which are in excellent agreement with experimental values.\cite{adam59} However, the estimated bulk modulus of 251~GPa is significantly smaller than the experimental value of 284~GPa.\cite{desgreniers99}. Jiang {\em et al.}\cite{jiang2010} also studied the electronic structure of hafnia polymorphs within the LDA approximation as well as within the $G_0W_0$ and $GW_0$ many-body perturbation schemes with good results for the latter for t-HfO$_2$.

The problem in obtaining quantitatively accurate results from such calculations stems from the fact that Zr and Hf have 4d and 5d electrons. In oxides, 3d, 4d, and 5d electrons are rather localized and lead to electronic correlation effects that are very difficult to capture using DFT. Hybrid functionals can often do a better job for semiconductors than more standard GGA or so called DFT+U calculations, in which a Hubbard $U$-parameter is added to the d-orbitals in addition to the regular GGA (or LDA) functional. However, there is no guarantee that hybrid functionals will give quantitatively more accurate results for transition metal oxides than do other exchange-correlation functionals\cite{perdew2014}. In addition, hybrid functionals are much more computationally expensive than local or semi-local functionals.

It is obviously of great interest to establish computational benchmarks both to guide DFT-based developments and modeling, but also to help guide experimental measurements where obtained results have great uncertainties. This is for example the case with measurements of the bulk modulus of zirconia polymorphs, especially t-ZrO$_2$ and c-ZrO$_2$, with the latter having experimental uncertainties of almost 100\%.\cite{ricca2015,aldebert85} In this work, we have performed quantum Monte Carlo (QMC)  simulations in order to establish such benchmarks and also to help establish properties with better certainty where experimental uncertainty is quite large, \emph{e.g.,} the bulk modulus of c-ZrO$_2$. Of course, the computational results are for T=0~K. We will therefore discuss our results from two different perspectives. The first one is just to compare and discuss our DFT and DMC calculations with other calculated values in the literature - such comparisons are unambiguous insofar as all calculations are at 0~K. The second perspective is to compare with experimental data. One possible way to do so is to add the effects finite temperatures to the calculations using the quasi-harmonic approximation to add entropic contributions from lattice vibrations. However, such calculations cannot presently be done within QMC. Therefore, the finite-temperature contributions would have to be done using DFT, and would then add to the QMC 0~K electronic energy the same DFT-based contribution at finite temperatures as it would to some other DFT calculations. That makes such comparisons rather meaningless since one would add the same finite-temperature contributions to DMC and DFT and so still end up comparing 0~K electronic energies. Instead, we will focus on using available experimental data for thermal expansion to extrapolate lattice parameters to 0~K for comparison with calculated lattice parameters\cite{stefanovich1994,milman2009,aldebert85}. This works rather well for the monoclinic and tetragonal phases, for which there exist rather extensive measurements, but is a bit more difficult for the cubic phases, for which there are fewer experimental data. There are also experimental data on bulk moduli and for some values of cohesive energies\cite{ackermann1975}. We will briefly discuss these from the perspective of first-principle modeling.

We also examine the energetics of tetragonal distortions of oxygen columns in t-ZrO$_2$ and t-HfO$_2$. In t-ZrO$_2$, these distortions play an important role in the tetragonal to cubic transition, and also greatly affect the optical gap (at the $\Gamma$ point) in t-HfO$_2$\cite{jiang2010,carbogno2014}. We compare the energetics of such distortions obtained from DFT with the energetics from DMC. Because of the computational expense of the DMC calculations, we had to restrict the motion of the oxygen columns to be strictly along the $c$-axis. Nevertheless, a comparison between the DFT and DMC results is instructive. We did also perform full structural optimization, including lattice vectors, for the tetragonal structure using the PBE0 hybrid functional to confirm that the distortion of the oxygen columns along the $c$-axis is the only relevant degree of freedom in the tetragonal distortion.

\section{\label{sec:level2}QMC methods}
\subsection{Variational and diffusion Monte Carlo\label{subsec:VDMC}}
The properties and behavior of materials and molecules can be accessed at the quantum level through solving the time-independent Schr\"odinger equation, $H\Psi (R)=E\Psi(R)$, where  $H$ is the Hamiltonian describing the interactions between the $N$ electrons at coordinates $r_1,\ldots,r_N$ and the atomic nuclei, $\Psi(r_1,...,r_N) $ is the many-body wavefunction, and $E$ the energy. 
The Hamiltonian describing electrons in a solid, is
 \begin{equation}\label{eq:ham}
 H = \sum_{i=1}^N  [ -\frac{1}{2}\nabla_i^2  + v_{ext} (\mathbf{r}_i) ]  +\sum_{i<j}^N \frac{1}{r_{ij}},
 \end{equation} 
where the first two terms are the kinetic energy and the external single-particle potential from the nuclei, respectively, and the last term describes the electron-electron interactions. This last term is responsible for moving the Schr\"odinger equation from a 3D partial differential equation to a 3$N$-dimensional partial differential equation, where $N$ is the number of electrons. The goal becomes to find the lowest eigenvalue in the 3$N$ dimensional space of the electrons. Many methods 
tend to reduce the complexity by approximating the last term. The approach in DFT is to study the equations of an auxiliary system of $N$ non-interacting electrons in a potential constructed to give the same ground state density as the system of interacting electrons.~\cite{Hohenberg1964,Kohn1965} The ground state energy can then be evaluated, provided the exact energy functional of the density is known. The great advantage of DFT is that the complexity of the problem is reduced to that of $N$ non-interacting electrons, but the accuracy depends on the accuracy of the approximate energy functional. DFT is considered the workhorse of electronic structure methods, and has proven to be very robust\cite{DFT_Science2016}, but at the same time the accuracy of the method is strongly dependent on the choice of the approximation.\\

In contrast, real-space QMC focuses on the many-body electronic wavefunction, but instead of performing integrals by explicit quadrature, one samples the many-body wavefunction by performing a random walk in the 3$N$-dimensional coordinate space of the electrons. Variational Monte Carlo (VMC) is a direct application of the Rayleigh-Ritz variational method.
If the trial wavefunction $\Psi_T(\{\mathbf{R}\})$, where $\{\mathbf{R}\}$ denotes the collection of $3N$ electron coordinates, satisfies the requisite symmetries and boundary conditions, then the ratio $E_V= \int d\{\mathbf{R}\} \Psi_T^*(\{\mathbf{R}\}) H \Psi_T(\{\mathbf{R}\}) / \int\,d\{\mathbf{R}\} |\Psi_T(\{\mathbf{R}\})|^2 $ is an upper bound to the exact ground state energy. 
In this work, we consider only the commonly-used generalized Slater-Jastrow function:
 \begin{equation}\label{eq:psi}
 \Psi_T(\{\mathbf{R}\}) = \exp\left[- J(\{\mathbf{R}\}) \right] \sum_k c_k {\det}_k (\phi_i (\mathbf{r}_j)),
 \end{equation}
where the Jastrow function $J(\{\mathbf{R}\})$ is a real function of all the electron coordinates, and is symmetric under electron exchange; it serves to correlate the electrons. The determinants keep the wavefunction antisymmetric, and the 3D functions $\phi_i$ in the determinants are the orbitals selected in some suitable way. The Jastrow function $J(\{\mathbf{R}\})$ and the orbitals are parameterized and their values optimized to minimize $E_V$. By avoiding the need to use forms that facilitate numerical quadrature, Monte Carlo sampling allows the use of much more complex trial wavefunctions encapsulating the electronic correlations. This is essential in order to describe the physics of the system, and most electronic structure methods lack accurate descriptions of correlations. 

VMC gives the lowest upper bound of the ground state energy consistent with the assumed trial wavefunction in Eq.~(\ref{eq:psi}).
Diffusion Monte Carlo (DMC) goes further by finding the lowest energy consistent with the nodes (or phase if it is complex) of the assumed trial function; only the nodal surface of the trial wavefunction is preserved to maintain the required anti-symmetry of the wavefunction. 
We define the many-body phase of the trial function by $\Theta (\{\mathbf{R}\}) = Im\lbrace \ln (\left[\Psi_T (\{\mathbf{R}\})\right] \rbrace$. Then in the class of functions $\exp[-J(\{\mathbf{R}\}) + i \Theta (\{\mathbf{R}\}) ]$, we minimize the ground state energy with respect to $J(\{\mathbf{R}\})$.  
This can be mapped to a random walk problem where instead of working in a space of a single system, it is formulated in an ensemble $\{R_i\}$ of $ P$ ``walkers''. Each walker executes a VMC random walk for a certain number of steps, then based on the error of the trial wavefunction, branches by either dying or making one or more copy of itself. DMC has been found to lower the VMC error by roughly an order of magnitude.\cite{rev:qmcsolids}\\

We use QMCPACK for the QMC calculations, and Quantum Espresso 5.3.0 for all DFT calculations with kinetic energy cutoffs of 350~Ry (450~Ry) for calculations of Zr and zirconia (Hf and hafnia). DFT calculations using Quantum Espresso were also used to generate the trial wavefunctions. Cohesive energies for hcp Zr and Hf were calculated by subtracting the calculated energy per atom of the bulk system from energy of an isolated single atom. For oxides, we calculated the energy per atom of the bulk oxide and subtracted the energy of a single isolated metal (Zr or Hf) atom and that of two isolated oxygen atoms. We did also use FHI-aims\cite{blum2009,ren2012,marek2014} to do full structural relaxations of t-ZrO$_2$ and t-HfO$_2$ using the ``default light'' settings and numerical orbitals, $4\times4\times4$ k-point meshes, and the PBE0 hybrid functional\cite{adamo1999jchemphys}.

\subsection{Pseudopotentials, validation and verification\label{subsec:PP}}
DMC is very expensive because many walkers are needed to reduce statistical noise. On the other hand, because the walkers are statistically independent, DMC codes scale superbly on leadership-class computers.\cite{Kim2012,Esler2012} However, even on leadership computers it is prohibitively expensive to include all electrons in calculations of, {\emph e.g.,} transition metal oxides; instead, pseudopotentials are used. In our work, scalar-relativistic pseudopotentials for zirconium and hafnium were generated with a plane-wave basis set, as implemented in OPIUM package.\cite{opium} The pseudopotentials were created using the local density approximation (LDA) exchange-correlation functional of DFT, with semi-core states included into valence. The zirconium (hafnium) pseudopotential was based on 28 (60) core electrons and 12 (12) valence electrons, and the oxygen pseudopotential was based on six valence electrons. The electronic configurations for the pseudopotentials were [Ar+3d$^{10}$]4s$^2$4p$^6$4d$^2$5s$^2$, [Pd+4f$^{14}$]5s$^2$5p$^6$5d$^2$6s$^2$ and [He]2s$^2$2p$^4$ for Zr, Hf, and O respectively. 
\begin{table}[t]
\centering
\caption{Estimated values of 1st-, 2nd ionization potential (IP), and electron affinity (EA) in units of eV for an isolated Zr and Hf atom.} 
\label{tab:DMC_atom_Zr}
\begin{tabular}{cccccc}
\hline \hline
     &     &      LDA    &  GGA &  DMC  &  Exp.   \\ \hline
 \multirow{2}{*}{Zr atom} & 1st IP &   5.35 &  5.03  & 6.43(2) &  6.63$^1$   \\ 
  &   EA &   2.03  &   1.75   & 0.41(3) & 0.43(1)$^2$   \\ \hline 
  \multirow{2}{*}{Hf atom} & 1st IP & 5.98 &  5.72 & 6.78(2) & 5.83$^1$ \\
  &    2nd IP &  11.88 & 11.84 & 14.55(2) & 14.90$^1$ \\ \hline \hline
\end{tabular}
\begin{flushleft}
$^1$Ref.~\onlinecite{lide03}.\\
$^2$Ref.~\onlinecite{feigerle98}.
\end{flushleft}
\end{table}

\begin{figure}[b]
 \includegraphics[width=3.0in]{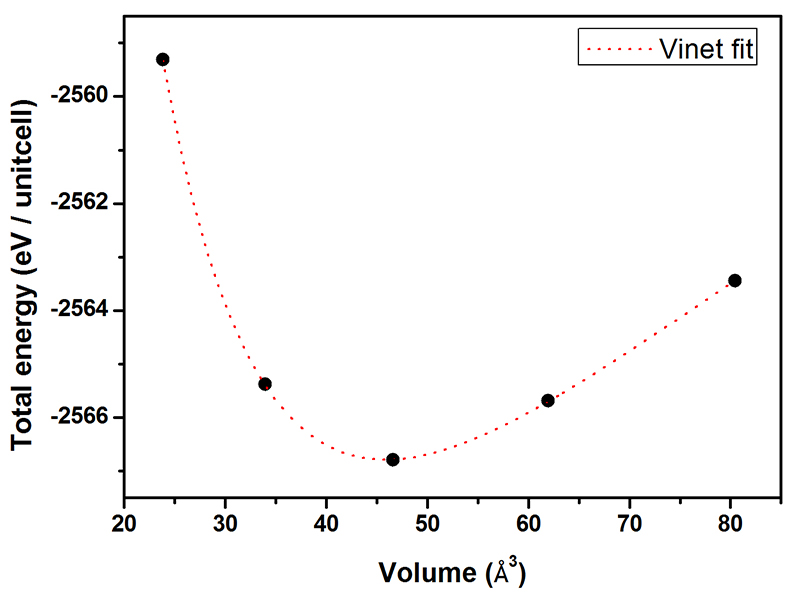}
 \caption{Total DMC energy for hcp Zr as function of unit cell volume. The Vinet fit to the calculated points is indicated.}
 \label{fig:Zr_EOS}
\end{figure}
\begin{table}[t]
\centering
\caption{Estimated values of the lattice constant ($a$), bulk modulus($B_0$), and cohesive energy ($E_{coh}$) for hcp zirconium and hafnium computed with DFT and DMC.} 
\label{tab:DMC_bulk}
\begin{tabular}{cccccc}
\hline \hline
   &       &    LDA   &  GGA  & DMC & Exp. \\ \hline
 \multirow{3}{*}{hcp Zr} &  $a$ (\AA) &   3.16       &   3.24$^1$    &  3.22(1) & 3.23$^1$ \\ 
&   $B_{0}$ (GPa) &  104       &    96$^1$    &  96(1) & 92(3)$^2$ \\
&    $E_{coh}$ (eV/atom) & 7.46 & - &  5.94(2) & 6.25$^3$  \\ \hline 
 \multirow{3}{*}{hcp Hf} & $a$ (\AA) &   3.13       &   3.19    &  3.20(1) & 3.19$^4$ \\ 
&   $B_{0}$ (GPa) &  119       &    108    &  110(1) & 110$^5$ \\
&   $E_{coh}$ (eV/atom) & 7.56 & 6.47 &  6.51(2) & 6.44$^3$  \\ \hline \hline
\end{tabular}
\begin{flushleft}
$^1$Ref.~\onlinecite{wang11}.\\
$^2$Ref.~\onlinecite{zhao05}.\\
$^3$Ref.~\onlinecite{kittel05}.\\
$^4$Ref.~\onlinecite{russell53}.\\
$^5$Ref.~\onlinecite{lide03}.\\
\end{flushleft}
\end{table}

In order to ascertain the accuracy of the pseudopotentials, we first calculated the atomic properties of the zirconium and hafnium atoms using DMC. 
To avoid interactions with spurious periodic image atoms stemming from the periodic boundary conditions in Quantum Espresso, we used a sufficiently large computational cubic cell of side 20~\text{\AA} with a single atom being located at the center of the cell.

The first and second ionization potential and electron affinity of an isolated atom can be calculated as $E(N-1)-E(N)$, $E(N-2)-E(N-1)$ and $E(N)-E(N+1)$, respectively. Here, $E(N-1)$, $E(N-2)$, $E(N+1)$, and $E(N)$ represent the total energy of charge-1 and charge-2 cations, anion, and neutral atom, respectively. 
We did not compute the DMC electron affinity for Hf because the electron configuration of an anion Hf, which is stabilized as the mixed state of 5$d^3$6$s$6$p$ and 5$d^2$6$p^3$, is difficult to fully implement in a Slater-Jastrow wavefunction, and instead we computed the first and second ionization potentials.
Table~\ref{tab:DMC_atom_Zr} 
shows the ionization potential and electron affinity for Zr, and first and second ionization potentials for Hf. The DMC results are in excellent agreement with corresponding experimental values, while LDA and GGA underestimate the atomic properties. 

We also calculated properties of bulk hcp zirconium and hafnium in order to verify the accuracy for the bulk systems of our DMC calculations with the pseudopotentials we used. For Zr, the initial DFT LDA trial wavefunctions were obtained using a $6\times6\times6$ $k$-point mesh, which provided $k$-point convergence. The DMC calculations used a supercell consisting of 16 Zr atoms (eight primitive unit cells). Twist-averaged boundary conditions with a total of 64 twists were applied to reduce one-body finite size effects.\cite{Ceperley:twists}    
In order to obtain the equation of state of bulk hcp zirconium, we computed the total energy as function of unit cell volume (Fig.~\ref{fig:Zr_EOS}), keeping $c/a$ fixed at the experimental value of 1.59 for Zr.
By using a Vinet fit,\cite{vinet86} we obtained equilibrium lattice constant, bulk modulus, and cohesive energy, as shown in Table~\ref{tab:DMC_bulk}. 
Even though we did not perform a two-body finite size analysis for Zr, the results for both lattice constant and bulk modulus are in good agreement with experimental values, and the DMC cohesive energy for Zr is calculated to be just 0.31(2)~eV/atom smaller than the experimental one. 

\begin{figure}[t]
 \includegraphics[width=3 in]{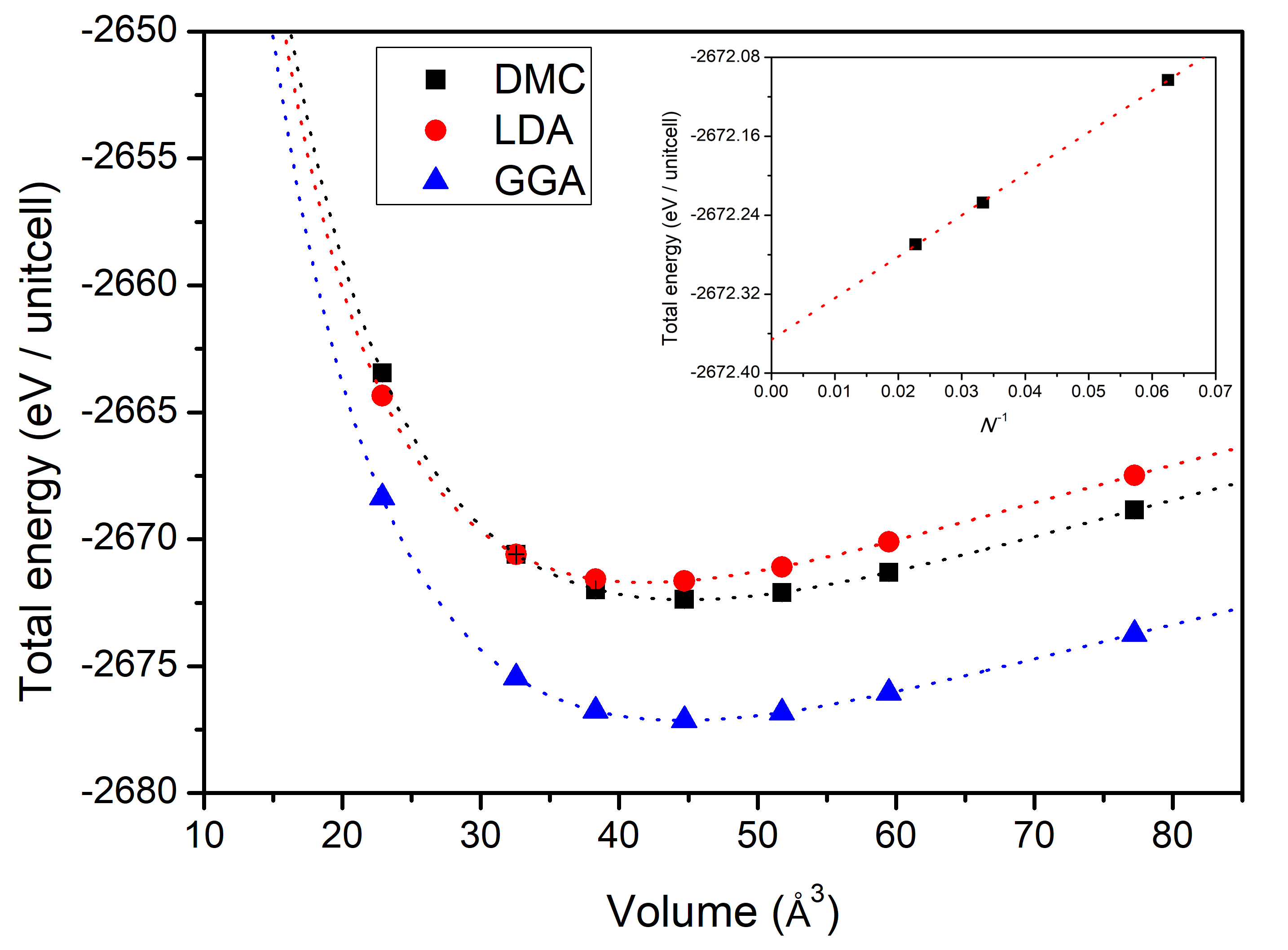}
 \caption{ DMC total energy of a hcp Hf as function of a unit cell volume. The dotted lines represent Vinet fits. (Inset) DMC total energy of hcp Hf as function of $1/N$, where $N$ is the total number of atoms in the computational supercell. The dotted line indicates a single linear regression fit.
 }
 \label{fig:EOS_bulk}
\end{figure}

Similarly, we computed DMC total energies for various volumes of a Hf hcp unit cell with the ratio of $c/a$ fixed at the experimental value of 1.58.\cite{russell53}
We generated trial wavefunctions with DFT calculations using LDA and GGA XC functionals on a $12\times12\times12$ $k$-point mesh for $k$-point convergence using the hcp primitive unit cell.
We also performed a finite-size analysis for the DMC result of bulk hcp Hf by twist-averaging DMC total energies over total 64 twists to reduce the one-body finite size effect.
We then extrapolated the twist-averaged DMC total energies computed at different hcp Hf supercells containing 8 (192 electrons), 15 (360 electrons), and 22 (528 electrons) atoms to the thermodynamic limit ($N\to\infty$) using a single linear regression fit, as shown in the inset of Fig.~\ref{fig:EOS_bulk}. 
Figure~\ref{fig:EOS_bulk} shows DMC and DFT (LDA and GGA) total energies as function of the Hf unit cell volume.  
From this EOS, the equilibrium lattice parameter, bulk modulus, and cohesive energy for a bulk Hf were obtained using a Vinet fit [see dotted lines in Fig.~\ref{fig:EOS_bulk}]. 
The results in Table~\ref{tab:DMC_bulk} show that DMC results for the structural parameters considered here are in good agreement with experimental values.
These results allow us to conclude that the Hf pseudopotential is appropriate to describe physical and electronic properties of a bulk Hf system, as well as the Hf atomic properties, within the QMC framework.

\section{\label{sec:results}Results and Discussion}
\begin{figure}[t]
 \includegraphics[width=3.0in]{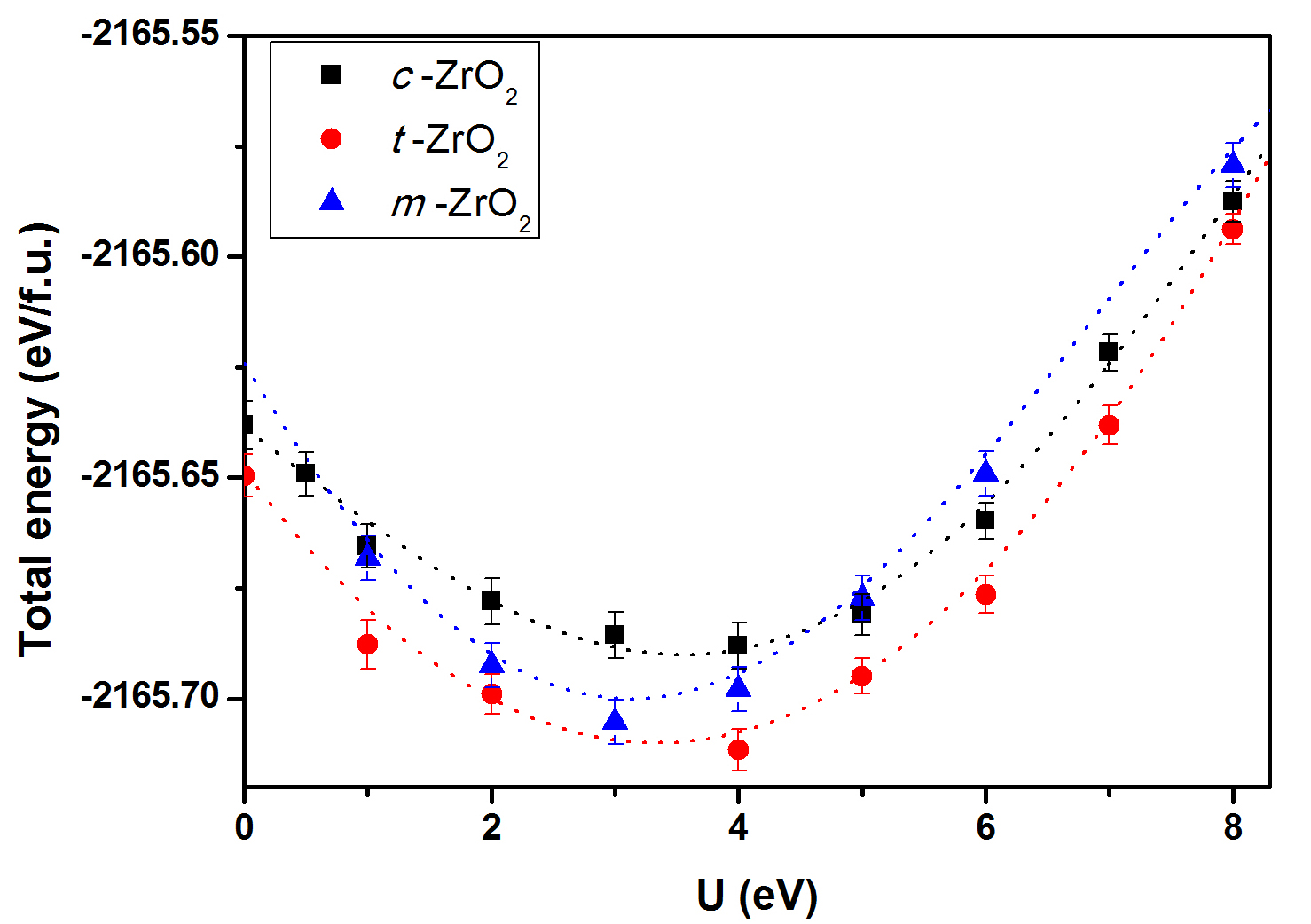}
 \caption{Total DMC energy for m-ZrO$_2$, t-ZrO$_2$, and c-ZrO$_2$ as function of Hubbard $U$ value in the initial trial wavefunction obtained from DFT using LDA+U.}
 \label{fig:DMC_U}
\end{figure}

In order to generate single-particle orbitals for trial wavefunctions, we used DFT within the LDA+U scheme, where the Hubbard $U$-parameter is treated as a variational parameter to optimize the DMC nodal surface of the wavefunction. The basis for this is the fact that DMC satisfies a strict variational principle. Therefore, by varying $U$ and calculating the DMC ground state energy as function of $U$, we can find an optimal nodal surface that minimizes the DMC ground state energy within the subspace spanned by the LDA+U trial wavefunctions. This has in practice proven to be an efficient way to generate DMC ground state properties in excellent agreement with experimentally accessible values\cite{BenaliTi4O7,shin17,shin17NiO}. We computed the DMC total energy of cubic, tetragonal, and monoclinic phases of ZrO$_2$ and HfO$_2$ as a function of $U$ in the trial wavefunction using experimental lattice parameters (see Table~\ref{tab:ZrO2_EOS}). Using a quartic fit, (see Fig.~\ref{fig:DMC_U}) the optimal values of $U$ for $c$-ZrO$_2$, $t$-ZrO$_2$, and $m$-ZrO$_2$ are calculated to be 3.54(12), 3.34(14), and 3.13(8)~eV, respectively. 

\begin{figure}[t]
\includegraphics[width=3.0in]{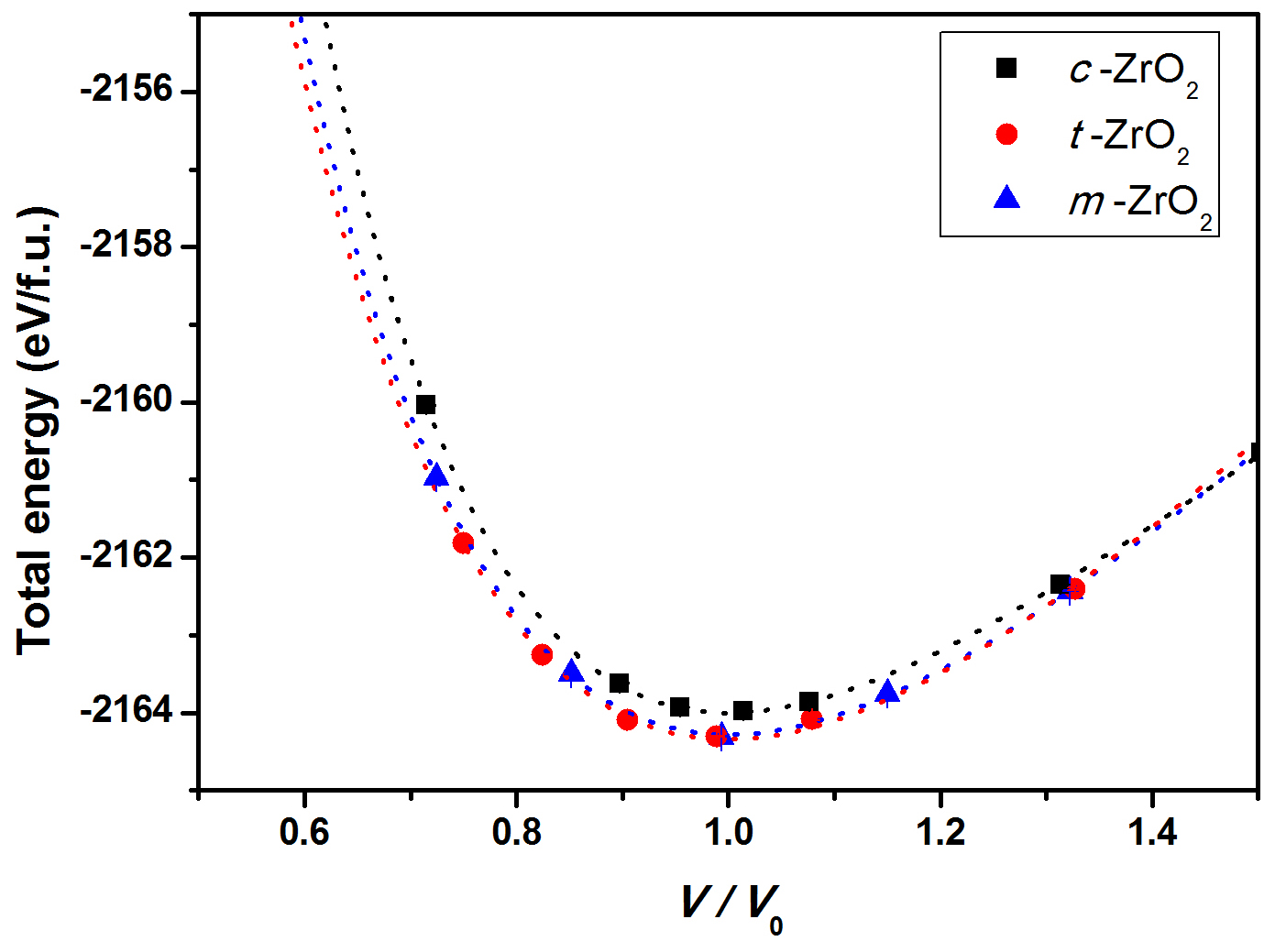}
 \caption{Total DMC energy for m-ZrO$_2$, t-ZrO$_2$, and c-ZrO$_2$ as function of volume ratio.}
 \label{fig:zr_phase}
\end{figure}

Using these values for $U$ to generate DFT trial wavefunctions, we then calculated the DMC energies for the $c$-ZrO$_2$, $t$-ZrO$_2$, and $m$-ZrO$_2$ structures as function of lattice constant, as shown in Fig.~\ref{fig:zr_phase}. For $m$-ZrO$_2$ ($t$-ZrO$_2$), the ratios between $a$, $b$, and $c$ ($a$ and $c$) axes were fixed while the volume of the unit cell was varied. For t-ZrO$_2$ the ratio was fixed at $c/(\sqrt{2}a)=1.02$. Equations of state were obtained using Vinet fits to the calculated data. 

\begin{figure}[t]
 \includegraphics[width=3 in]{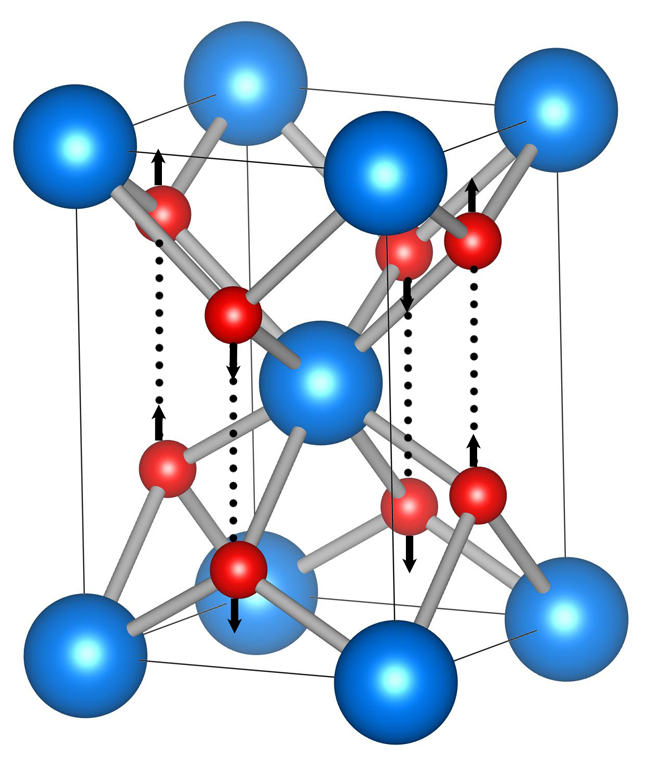}
 \caption{Directions of the oxygen pair distortion in the tetragonal phases of ZrO$_2$ and HfO$_2$. The blue and red spheres indicate Zr, or Hf, and O atoms, respectively.} 
 \label{fig:dz}
\end{figure}

\begin{figure}[t]
 \includegraphics[width=3.0in]{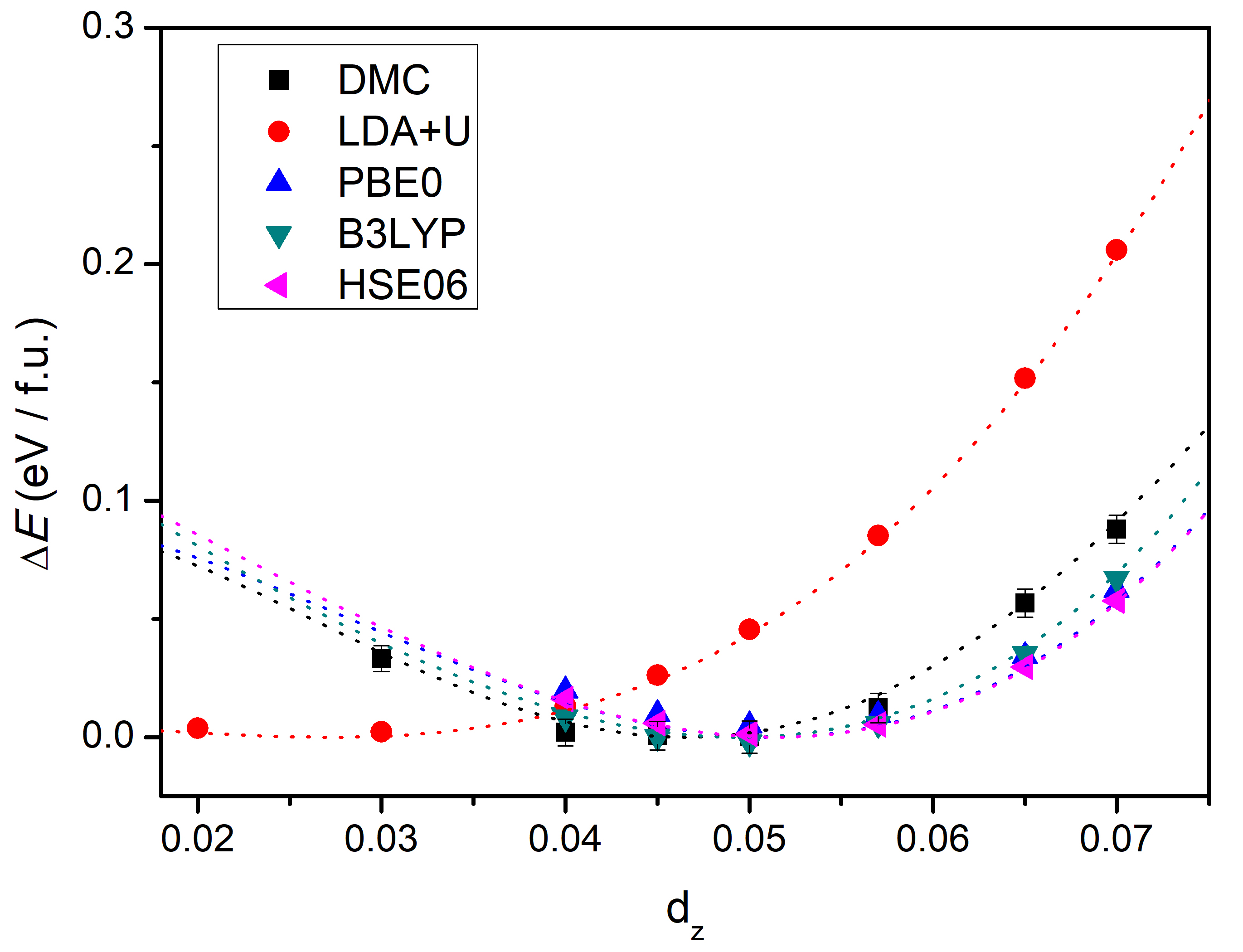}
 \caption{Total energy for t-ZrO$_2$ as function of $d_z$ away from its equilibrium value. The dotted line represents quartic fits. The equilibrium DMC value of $d_z$ is estimated to be 0.047(1), which is in excellent agreement with experimental value of 0.047.\cite{bouvier01}}
 \label{fig:ZrO2_dz}
\end{figure}

In t-ZrO$_2$ (as well as in t-HfO$_2$) an additional degree of freedom is the distortion of pairs of oxygen atoms located in columns in the tetragonal unit cell\cite{fabris2001,sternik2005,carbogno2014}. Carbogno {\em et al.}\cite{carbogno2014} showed, using the HSE06 hybrid functional, that displacements of the oxygen atoms in the tetragonal basal plane are important for the tetragonal to cubic transformation. Because the thermodynamics of this transition is beyond the scope of our work, and because of the great expense in the DMC calculations, we restricted the displacements to be along the $c$-direction (see Fig.~\ref{fig:dz}). This excludes possibilities of obtaining the minimum energy barrier for distortion of the oxygen atoms\cite{carbogno2014} but does allow us to find the distortion at which the t-ZrO$_2$ structure has minimum energy.
We calculated the total energy, both DMC as well as for a variety of DFT schemes, of t-ZrO$_2$ as a function of a distortion parameter $d_{z}$, defined as $z_{O}=(0.25 \pm d_{z})c $ where $z_{O}$ is the location of the oxgyen atoms along the $c$-direction. 
Note that in these calculations we kept the lattice parameters $a$ and $c$ fixed at their extrapolated or measured low-temperature experimental values\cite{aldebert85,stefanovich1994} of 
3.57~{\AA} and 5.18~{\AA} while varying $d_{z}$.
Using a quartic fit, we found the equilibrium distortion of the oxygen pairs for DMC as well as for DFT using a variety of XC functionals, as shown in Fig.~\ref{fig:ZrO2_dz}(b).
The minimum value of the DMC total energy is attained at $d_{z}=0.047(1)$, which is in excellent agreement with experimental value of 0.047.\cite{bouvier01} For t-ZrO$_2$, the location of the minimum energy is rather insensitive to which DFT hybrid functional is used. 
\begin{table*}[t]
\centering
\caption{Fully optimized lattice parameters and $d_{z}$ for t-ZrO$_{2}$ and t-HfO$_{2}$ using FHI-aims (FHI-aims) and Quantum  Espresso [Q.E. (fr)].
Note that only atomic coordinates were optimized in Q.E. (fr), with fixed lattice parameters obtained from FHI-aims. 
``Q.E. (s)" indicates manually optimized d$_{z}$ values using fixed experimental lattice parameters in Tables~\ref{tab:ZrO2_EOS} and~\ref{tab:HfO2_EOS}, and fixed Zr atomic coordinates.}
\label{tab:dz}
\begin{tabular}{c|ccc|ccc}
\hline\hline
  & \multicolumn{3}{c|}{t-ZrO$_2$} & \multicolumn{3}{c}{t-HfO$_2$} \\ 
PBE0   &    $a$ (\AA)      &   $c$ (\AA)   & $d_{z}$       &    $a$ (\AA)  &  $c$ (\AA) &  $d_{z}$ \\ \hline
FHI-aims  &  3.59 & 5.19 & 0.054 & 3.58 & 5.17 & 0.050 \\
Q.E. (fr)  &  3.59 & 5.19 & 0.051 & 3.58 & 5.17 & 0.057 \\
Q.E. (s) & 3.57 (fixed) & 5.18 (fixed) & 0.051 & 3.64 (fixed) & 5.29 (fixed) & 0.069 \\ \hline\hline
\end{tabular}
\end{table*}

As mentioned earlier in the Methods section, we did a full optimization of the six-atom t-ZrO$_2$ unit cell, including relaxing the lattice vectors using FHI-aims\cite{blum2009,ren2012,marek2014} with the PBE0 hybrid functional\cite{adamo1999jchemphys} and the default ``light'' setting until the maximum force on any atom was less than 0.006~eV/{\AA}. We then used the FHI-aims optimized lattice vectors as input to a QE relaxation with PBE0 of the internal atomic coordinates, keeping the lattice vectors fixed. In addition, we performed a manual optimization of the oxygen distortion using QE and PBE0 in which we used fixed experimental lattice parameters from Tables~~\ref{tab:ZrO2_EOS} and~\ref{tab:HfO2_EOS} and kept the Zr positions fixed. There is some difference between the optimized $d_z$-values using FHI-aims and QE using the same lattice vectors. We attribute this difference to 
smaller basis set used in FHI-aims in these calculations compared to the very large kinetic energy cutoff in QE. Taking these differences into account, we take these calculations to confirm that the motion of the oxygen columns along the $c$-axis is the only relevant internal degree of freedom, justifying the one-parameter sweep of $d_z$ discussed earlier. Note the larger variation in $d_z$ for t-HfO$_2$, especially when the finite-temperature experimental lattice parameters are used (to be discussed later).



\subsection{Zirconia}
\subsubsection{Zirconia -- comparisons between computed results}
Our main results for the zirconia structural parameters are summarized in Table~\ref{tab:ZrO2_EOS}, both for DMC as well as for various DFT schemes and functionals, including representative recent values from the literature. 
In general, the hybrid functionals PBE0 and HSE06 give lattice parameters that are in good agreement with each other as well as with the DMC values. The results by Ricca et al.\cite{ricca2015} are an exception. We suspect the origin of this is the basis set used in that work, given that the FHI-aims and QE calculations generally agree very well with each other. As is usually the case, LDA tends to overbind and GGA (eg. PBE) underbind, with too small and too large lattice parameters, respectively. It is interesting to note that LDA+U corrects (increases) the lattice parameters compared to LDA but then substantially reduces the bulk modulus while LDA gets the bulk modulus right. In contrast, all other DFT flavors underestimate the bulk modulus, especially for the cubic and tetragonal phases. 
All DFT calculations by us and those quoted here (as well as many others), except for one, get the energetic ordering correctly, with m-ZrO$_2$ having the lowest energy, followed by t-ZrO$_2$ and c-ZrO$_2$ in increasing energy. The exception is the calculation by Ricca et al., which gets the ordering of the tetragonal and cubic phases wrong. Again, we suspect this error stems from the basis set used in that calculation. 
We note that LDA+U gets very good agreement for the cohesive energy compared to DMC, while the other DFT flavors underestimate the cohesive energy for the three phases. This clearly suggests that the shape and self-interaction corrections of the 4d orbitals play a central and sensitive role in getting both geometry and energetics right. LDA+U improves both geometry and cohesive energy relative to LDA, meaning that the location of the energy minima for the three phases in configuration space are improved by adding a Hubbard-U on the 4d orbitals, but the steepness of the potential wells about the minima then becomes much too small. The hybrid functionals, which include a range-dependent exact exchange energy, end up somewhere in between LDA+U and DMC: the lattice parameters are good, but the cohesive energies and bulk moduli are too small. It should also be pointed out that the properties of the LDA+U calculations depend very much on the values of the Hubbard U; ours were selected by minimizing the DMC total energy which serendipitously gave good values for lattice parameters and cohesive energy. Using another scheme to select U, e.g., by adjusting U to fit the optical band gap, would no doubt give different results for geometry and energetics.
%
Finally, we note that the energy differences between the different phases are very small. Therefore, small errors in cohesive energies will give large relative errors in energy differences. The DMC energy difference between the monoclinic and tetragonal phases is larger, 0.07~eV, than the uncertainty in the DMC energy difference, 0.04~eV. In contrast, the energy difference between the tetragonal and cubic phases is much smaller, 0.04~eV, and barely larger than the uncertainty, 0.03~eV. 

\subsubsection{Zirconia -- comparison with experiments.}
Table~\ref{tab:ZrO2_EOS2} shows the results from our HSE06 and DMC calculations together with experimental data. As can be expected, the agreement is generally best for the low-temperature monoclinic phase, for which thermal motion would have the smallest effect. In addition to the lattice parameters, the DMC cohesive energy for m-ZrO$_2$ is also in good agreement with the experimental value, while the experimental bulk modulus is much smaller than the calculated one; we note, however, the large experimental uncertainty in the measurement. Also, we suspect that even at room temperature the bulk modulus is softened by thermal motion, while presumably relatively small anharmonicity in the thermally induced lattice vibrations leads to no significant increase of the lattice parameters.

The agreement between the calculated lattice parameters and the measured ones is also quite good for the tetragonal phase. Reference~\onlinecite{stefanovich1994} presented lattice parameters for t-ZrO$_2$ extrapolated to zero temperature based on the thermal expansion measured in Ref.~\onlinecite{aldebert85}, and Refs.~\onlinecite{bouvier00,bouvier01} presented data from room- or low-temperature measurements of nano-crystalline t-ZrO$_2$. Therefore, in both of these measurements, the effects of thermal expansion on the lattice parameters are eliminated or are negligible. The tetragonal distortion $d_z$ measured at low temperatures\cite{bouvier01} is also in good agreement with the DMC value. However, as can be expected, the measured bulk modulus is much lower than the calculated ones, even room-temperature measurements (up to 200~GPa) of nano-crystalline t-ZrO$_2$\cite{bouvier00}; thermal motion significantly softens the bulk modulus.

Finally, the calculated lattice constant for the cubic phase is smaller than the measured one, 5.27~{\AA}\cite{aldebert85} at high temperatures (2,410$^\circ$~C) - here thermal expansion clearly plays a role. The extrapolated lattice constant in Ref.~\onlinecite{stefanovich1994} as well as our own extrapolation of the lattice constants using thermal expansion data (Fig.~\ref{fig:expansion}) give a lattice constant of 5.09~{\AA}, in very good agreement with the calculated one. We caution that because of the few available data the result - and its very good agreement with DMC - based on our simple linear fit may be accidental. However, our own similar extrapolations for the tetragonal and monoclinic phases, for which there are more experimental data, yield very good agreement with DMC values as the actual measurements (Fig.~\ref{fig:expansion}. Finally, the measured high-temperature bulk modulus for the cubic phase is much smaller than the calculated ones, as is to be expected.

It is remarkable that the calculated DMC energy differences $\Delta E_{t,c}$ and $\Delta E_{m,t}$ agree rather well with the measured enthalpy differences in Ref.~\onlinecite{aldebert85}. This could either be fortuitous, or an indication that the entropic contributions to the enthalpy differences are substantially equal at the transition temperatures.

\begin{table*}[t]
\centering
\caption{Results for structural parameters, bulk modulus, and cohesive energy from DFT calculations together with DMC data. $\Delta E_{t,c}$ and $\Delta E_{m,t}$ are the cohesive energy differences between t-ZrO$_2$ and c-ZrO$_2$, and between m-ZrO$_2$ and t-ZrO$_2$, respectively.}
\label{tab:ZrO2_EOS}
\begin{tabular}{cccccccccccc}
\hline \hline
                         phase &      & LDA  & LDA+U & GGA  &PBE$^1$& PBE0 &PBE0$^{2}$ & B3LYP  & HSE06 & HSE06$^1$ & DMC \\ \hline
\multirow{3}{*}{c-ZrO$_2$}&  $a$ (\AA)& 5.03 & 5.09  & 5.10  &5.03  & 5.06 & 5.11       & 5.10  & 5.06  & 5.08  & 5.07(1)  \\
                        &$B_0$ (GPa)  & 279  & 247   & 242   &      & 270  &            & 254   & 269   &       & 278(2)   \\ 
              &  $E_{coh}$ (eV/f.u.)  & 25.84 &23.36 & 22.11 &      &21.65 &            & 20.76 & 21.50 &       & 23.15(2) \\ \hline
\multirow{6}{*}{t-ZrO$_2$}&  $a$      & 3.56  & 3.58 & 3.61  &3.63  & 3.58 & 3.62       & 3.62  & 3.58  & 3.60 
                                                                                                                &3.58(1)  \\
                          &  $c$      & 5.16  & 5.19 & 5.24  &5.29  & 5.19 &5.245       & 5.25  &  5.19 & 5.20
                                                                                                               &5.19(2)  \\
                    & $c$/$\sqrt{2}a$ & 1.02  & 1.02 & 1.02  &1.03  & 1.02 &1.023       & 1.02  & 1.02  & 1.02 & 1.02 \\
                          &  $d_{z}$  & 0.048 & 0.027& 0.049 & 0.056&0.051 &0.052       & 0.049 & 0.051 & 0.051& 0.047(1) \\
                              & $B_0$ & 263   & 244  & 234   &      &259   &            & 244   & 258   &      &
                                                                                                                 265(1) \\ 
                         &  $E_{coh}$ & 26.20 & 23.51& 22.21 &      &21.73 &            & 20.86 & 21.60 &      & 23.19(2) \\ 
                  &  $\Delta E_{t,c}$ & 0.36  & 0.15 & 0.10  &0.14  & 0.08 &-0.025      & 0.10  & 0.10  & 0.08 &0.04(3) \\ \hline
\multirow{6}{*}{m-ZrO$_2$}  &  $a$    & 5.10  & 5.16 & 5.17 &       & 5.13 &5.18        & 5.17  & 5.13  &      &5.14(1) \\
                            &  $b$    & 5.16  & 5.22 & 5.23 &       & 5.19 &5.24        & 5.23  & 5.19  &      &5.20(1) \\
                            &  $c$    & 5.27  & 5.33 & 5.34 &       & 5.30 &5.32        & 5.34  & 5.30  &      &5.30(1) \\
                            & $B_0$   & 255   & 232  & 227  &       & 251  &            & 237   & 253   &      &254(1) \\
         
               &  $E_{coh}$           & 26.25 & 23.67& 22.33&       & 21.82 &           & 20.99 & 21.71 &      &23.26(2) \\ 
               &  $\Delta E_{m,t}$    & 0.05  &  0.16& 0.12 & 0.13  & 0.09  & 0.062     & 0.13  & 0.11  & 0.075&0.07(4) \\ \hline \hline
\end{tabular}
\begin{flushleft}
$^1$Ref.~\onlinecite{carbogno2014}.\\
$^2$Ref.~\onlinecite{ricca2015}.\\
\end{flushleft}
\end{table*}
\begin{table*}[t]
\centering
\caption{Results for structural parameters, bulk modulus, and cohesive energy from HSE06 and DMC together with available experimental data. $\Delta E_{t,c}$ and $\Delta E_{m,t}$ are the cohesive energy differences between t-ZrO$_2$ and c-ZrO$_2$, and between m-ZrO$_2$ and t-ZrO$_2$, respectively.}
\label{tab:ZrO2_EOS2}
\begin{tabular}{ccccccccccccc}
\hline \hline
                    phase &           & HSE06  & DMC  & Exp$^1$ & Exp.$^2$& Exp.$^3$&Exp.$^4$&Exp.$^5$& Exp.$^6$& Exp.$^7$&Exp.$^8$& Exp.$^9$ \\ \hline
\multirow{2}{*}{c-ZrO$_2$}&  $a$ (\AA) & 5.06  &5.07(1)& -      &-        & 5.09    & -      & -      & -       &    -    &     -  & -   \\
                   &$B_0$ (GPa)        & 269   & 278(2)& 248&-        & -       & -      & -      & -       &   -     & -      & - \\ \hline

\multirow{5}{*}{t-ZrO$_2$}&  $a$      & 3.58   &3.58(1)& -      &3.59     & 3.57    &-       & -      &         & -       & -      & -\\
                          &  $c$      & 5.19   &5.19(2)&        & 5.18    &   5.18  &-       &-       &-        &-        &-       & - \\
                    & $c$/$\sqrt{2}a$ & 1.02   & 1.02  & -      &1.02     & -       &-       & -      & -       & -       & -      & - \\
                           &  $d_{z}$ & 0.051  & 0.047(1)& 0.057& -       &-        &0.047   & -      &  -      & -       &  -     & - \\
                             & $B_0$  & 258    & 265(1)  & -    &  172(6) &-        & -      & 170-200& -       & -       & -      & - \\ 
                   &  $\Delta E_{t,c}$& 0.10 & 0.04(3)  &  -    &-        & -       &-       & -      & -       & -       &-       & 0.057 \\ \hline
\multirow{6}{*}{m-ZrO$_2$}  &  $a$    & 5.13 & 5.14(1)  & -     &  -      & -       &-       & -      &  5.15   & -       & -      & -\\
                            &  $b$    & 5.19 & 5.20(1)  & -     &-        & -       &-       & -      & 5.21    &-        & -      & - \\
                            &  $c$    & 5.30 & 5.30(1)  & -     & -       & -       &-       & -      & 5.31    & -       & -      & - \\
                            & $B_0$   & 253  & 254(1)   & -     &-        & -       & -      & -      &-        &212(24)  & -      & - \\ 
                       &  $E_{coh}$   & 21.71& 23.26(2) & -     &-        & -       & -      & -      & -       &         & 22.85  &- \\ 
                  &  $\Delta E_{m,t}$ & 0.11 & 0.07(4)  & -     & -       & -       & -      & -      & -       & -       & -      & 0.063\\ \hline \hline
\end{tabular}
\begin{flushleft}
$^1$Ref.~\onlinecite{kandil1984} room T measurements of yttrium-stabilized c-ZrO$_2$ extrapolated to 0\% yttrium.\\
$^{2}$Ref.~\onlinecite{bouvier00} at room T using nanocrystalline t-ZrO$_2$.\\
$^3$Ref.~\onlinecite{stefanovich1994} extrapolation to zero T using data in Ref.~\onlinecite{aldebert85}.\\
$^4$Ref.~\onlinecite{bouvier01} low-temperature measurement of nanocrystalline t-ZrO$_2$.\\
$^5$Ref.~\onlinecite{fukuhara93} used yttria-stabilized t-ZrO$_2$ at room T, Ref.~\onlinecite{bouvier03} used nanocrystalline t-ZrO$_2$ at room T.\\
$^6$Ref.~\onlinecite{howard88} at room T.\\
$^7$Ref.~\onlinecite{desgreniers99} at room T.\\
$^8$Ref.~\onlinecite{lynch74}.\\
$^9$Estimated in Ref.~\onlinecite{stefanovich1994} from measured enthalpy differences in pure t-ZrO$_2$ in Ref.~\onlinecite{ackermann1975}.\\
\end{flushleft}
\end{table*}

\subsection{Hafnia}
As with zirconia, we started by optimizing the DMC nodal surface by calculating the DMC ground state energy of the three hafnia polymorphs as a function of $U$ in the LDA+U DFT trial wavefunction. We used supercells consisting of four primitive unit cells with lattice parameters and geometries fixed at experimental values, and using a total of 64 twists (see Table~\ref{tab:HfO2_EOS}).
The self-consistent DFT LDA+U calculations for cubic, tetragonal, and monoclinic phases were done with $6\times6\times6$, $6\times6\times6$, and $4\times4\times4$ $k$-point meshes, respectively.
\begin{figure}[t]
 \includegraphics[width=3.0 in]{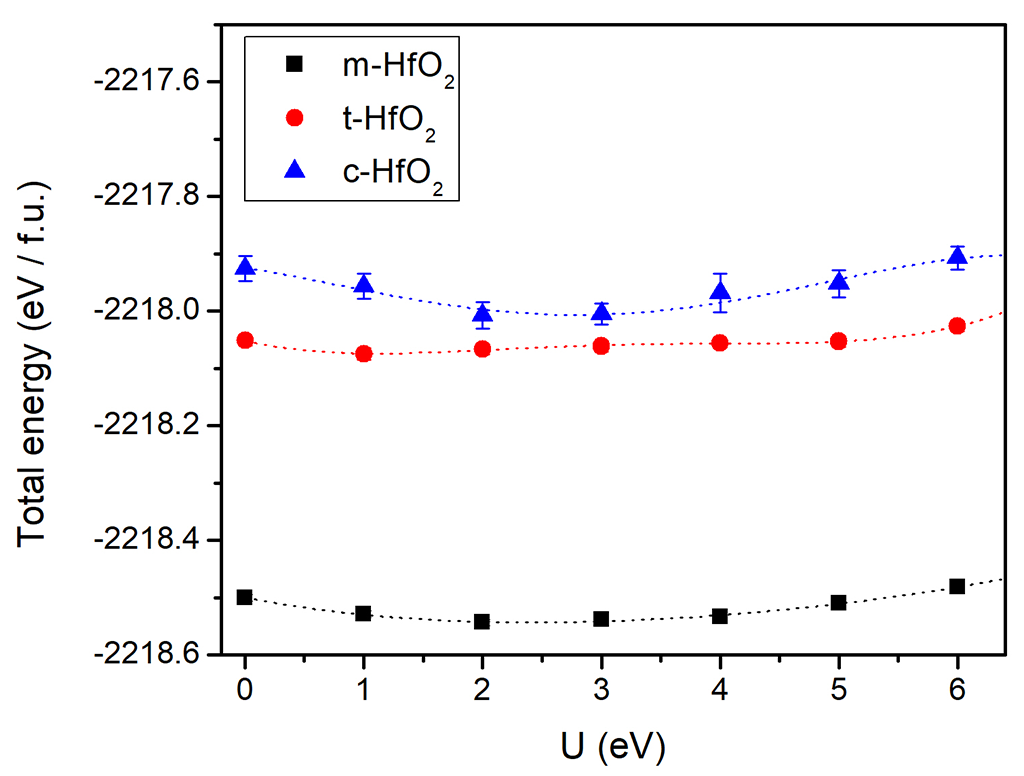}
 \caption{DMC total energy as function of U in the LDA trial wavefunction. Dotted lines indicate quartic fits.}
 \label{fig:HfO2_U}
\end{figure}

Figure~\ref{fig:HfO2_U} (a) depicts the DMC total energy as a function of the variational $U$-parameter using an LDA+U trial wavefunction for the hafnia polymorphs. 
The DMC ground state energy for t-HfO$_2$ behaves quite differently from that of the other hafnia polymorphs, as well that of ZrO$_2$ or TiO$_2$ polymorphs\cite{BenaliTi4O7,shin17}: it does not exhibit a pronounced minimum at the optimal value $U_{\rm opt}$ of $U$, but a shallow minimum at a small value of about 1~eV and a dimple that is almost a local minimum at a value of $U$ of about 6~eV. Also, in comparison of the DMC energy as function of $U$ for zirconia and hafnia polymorphs with that of antiferromagnetic NiO\cite{shin17NiO}, the the difference between the minimum DMC energy at $U_{\rm opt}$ and that at $U=0$ is very small, less than 0.1~eV/f.u. compared to about 0.7~eV/f.u. for NiO. This indicates that the on-site Coulomb repulsion is less significant in zirconia and hafnia compared to NiO, consistent with the fact that the on-site repulsion drives NiO to a Mott insulator antiferromagnetic ground state, while zirconia and hafnia do not have magnetic ground states.
Using a quartic fit for the DMC total energy as function of $U$ in the LDA+U trial  wavefunction, we find U$_{\text{opt}}$ to be  2.7(3), 1.1(3), and 2.4(2) eV for c-HfO$_2$, t-HfO$_2$, and m-HfO$_2$ , respectively.

\subsubsection{Hafnia - comparisons between computed results}
\begin{figure}[t]
 \includegraphics[width=3 in]{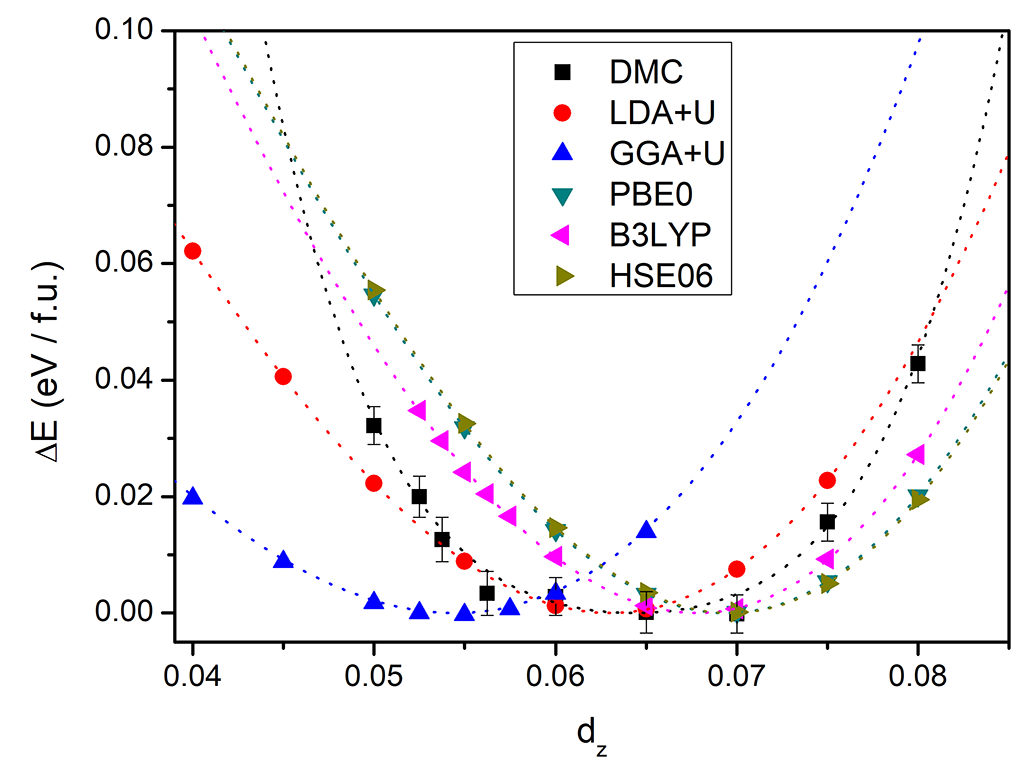}
 \caption{Total energy for t-HfO$_2$ as function of $d_z$ away from its equilibrium value. Here, the DMC total energies were computed in eight t-HfO$_2$ supercells with a total of 64 twists. Dotted lines represent quartic fits.}
 \label{fig:tHfO2_dz}
\end{figure}

We then continued to calculate the EOS and structural parameters for the HfO$_2$ polymorphs. In order to reduce structural degrees of freedom in the tetragonal and monoclinic phases, we imposed constraints on the lattice parameters while the unit cell volume was changed. 
The ratio $c/a$ in t-HfO$_2$ was fixed at its experimental value\cite{adams91} of 1.45; only a single degree of freedom, the lattice parameter $a$, was considered for m-HfO$_2$ because of the large (four) number of structural parameters that otherwise would have to be optimized. 
Therefore, $b/a$, $c/a$, and $\beta$ were fixed at their experimental values\cite{ruh70}.      For t-HfO$_2$, as was the case for t-ZrO$_2$, we calculated the total energy as function of oxygen distortion $d_z$ at fixed $c$ and $a$ at their experimental values of $a=3.64~\text{\AA}$ and $c=5.29~\text{\AA}$\cite{adams91}. Using a quartic fit, we found the equilibrium distortion for DMC and each of the DFT XC functionals [Fig.~\ref{fig:tHfO2_dz}(b)].
The minimum value of the DMC total energy is at $d_{z}=0.064(1)$, which is larger value than experimental value of $d_{z} = 0.047$ for tetragonal zirconia measured at room T.\cite{aldebert85,howard88,stefanovich1994}
We note that a previously obtained value of $d_z=0.05$  from DFT using the Perdew-Wang parametrization of the GGA XC functional is significantly smaller than both our DMC and DFT results, including GGA with the PBE parametrization.\cite{terki08} 
We attribute this discrepancy to a sensitive dependence of the structural optimization on the DFT XC functional and pseudopotentials (the DMC pseudopotentials are much ``harder" than pseudopotentials commonly used in DFT).
Figure~\ref{fig:tHfO2_dz}(b) shows that the energy difference $\Delta E$ at previously reported\cite{foster02,Iskandarova03,terki08} DFT equilibrium values of $d_{z}\sim 0.050$ - $0.07$ is well below 0.08~eV/f.u. for all of our DFT and DMC calculations.   
Therefore, the changes in total energy as the oxygen distortion is varied around $d_{z} = 0.05$ - $0.07$ are very small, which means that optimization of $d_{z}$ is sensitive to the choice of XC functional or pseudopotentials because of the very small contribution of $d_{z}$-minimization to the total energy of t-HfO$_2$. This is also evident in the results in Table~\ref{tab:dz}, where the obtained optimal value of $d_z$ is much more sensitive to choice of method (full potential and localized orbitals vs. pseudopotential and plane waves) as well as to lattice parameters. As is also seen in Table~\ref{tab:dz}, with the FHI-optimized lattice parameters of $a$=3.58~{\AA} and $c$=5.17~{\AA}, the calculated $d_z$ is much smaller, 0.05~{\AA} to 0.057~{\AA}, much more similar to the corresponding values for t-ZrO$_2$.
\begin{figure}[t]
 \includegraphics[width=3 in]{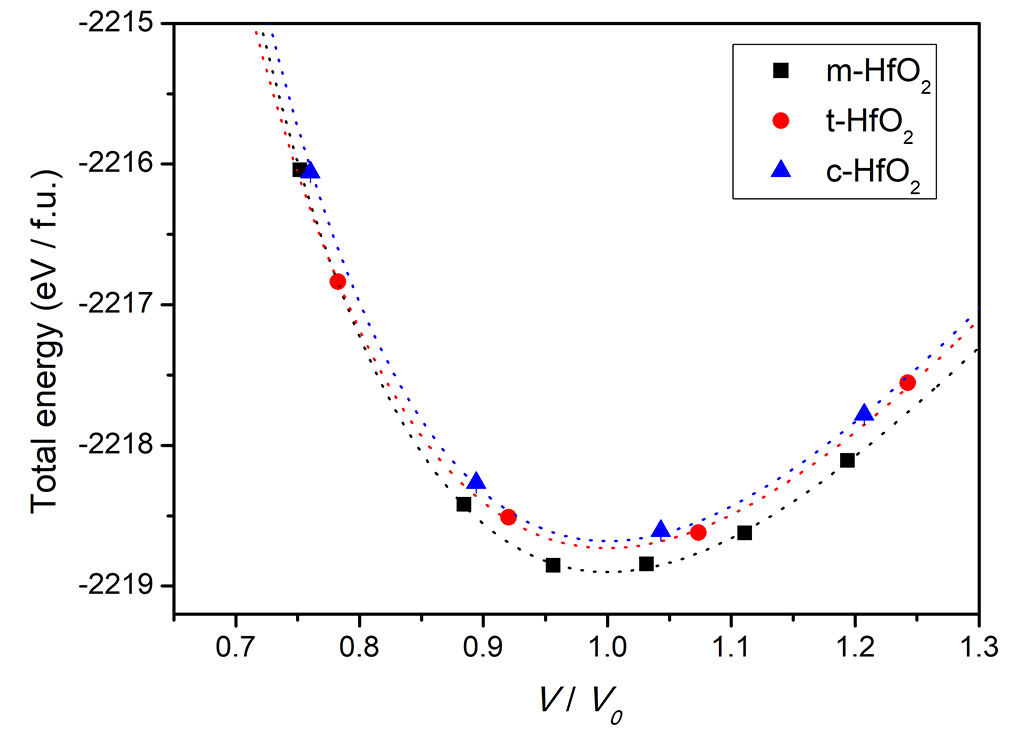}
 \caption{DMC total energy of a monoclinic, tetragonal, and cubic phase of hafnia as function volume ratio. $V$ is the unit cell volume that is varied, and $V_0$ is the equilibrium volume obtained from a Vinet fit. The dotted lines are Vinet fits.}
 \label{fig:EOS_HfO2}
\end{figure}
For t-HfO$_2$ the ratio of $c$ to $a$ was fixed at $c/(\sqrt{2}a)=1.03$, and $d_z$ fixed at its optimal value as a fraction of $c$ when calculating total energy. Equations of state were then obtained using Vinet fits. 
We considered a series of supercells (twists) in order to extrapolate DMC twist-averaged total energies to the thermodynamic limit: 6(64), 22(27), and 27(27) for c-HfO$_2$; 8(64), 16(64), and 22(27) for t-HfO$_2$; and 4(64), 6(64), 8(27) for m-HfO$_2$.

\begin{table*}[t]
\centering
\caption{Results for structural parameters, bulk modulus, and cohesive energy for hafnia polymorphs from DFT calculations together with DMC data. $\Delta E_{t,c}$ and $\Delta E_{m,t}$ represents cohesive energy difference between t-HfO$_2$ and c-HfO$_2$, and that between m-HfO$_2$ and t-HfO$_2$, respectively.}
\label{tab:HfO2_EOS}
\begin{tabular}{ccccccccccccc}
\hline \hline
 phase &                                    & LDA & LDA+U &LDA+U$^1$ &GGA &PBE$^2$& GGA+U & PBE0 & B3LYP & HSE06& HSE$^3$ & DMC \\ \hline
  \multirow{3}{*}{c-HfO$_2$}  &  $a$ (\AA)  & 4.97 & 5.01 &-& 5.03 &5.08& 5.07 & 5.00 & 5.04 & 5.00 &-& 5.04(1)  \\
                              & $B_0$ (GPa) & 299   & 273  &- & 268 &258& 238 & 287 & 271 & 286 &-& 297(1) \\ 
                    &  $E_{coh}$ (eV/f.u.)  & 26.83 & 24.73&- & 22.97 &-& 20.76 & 22.65 & 21.79 & 22.61 &-& 24.32(3) \\ \hline
  \multirow{6}{*}{t-HfO$_2$}  &  $a$        & 3.54 & 3.54 &-  & 3.58 &3.59& 3.60 & 3.56 & 3.59 & 3.57 &-& 3.58(2) \\
                              &  $c$        & 5.14 & 5.14 &-  & 5.20 &5.23& 5.23 & 5.17 & 5.22 & 5.19 &-& 5.20(3) \\
                    & $c$/$\sqrt{2}a$       & 1.03 & 1.03 & - &1.03 &1.03& 1.03 & 1.03 & 1.03 & 1.03 &-& 1.03 \\
                              &  $d_{z}$    & 0.067 &0.063&-  & 0.068 &0.055& 0.054 & 0.069 & 0.068 & 0.070 &-& 0.064(1) \\
                              & $B_0$       & 273 & 265   &-  & 245  &244& 226 & 267 & 253 & 267 &-& 271(9) \\ 
                         &  $E_{coh}$       & 26.93& 25.96&-  & 23.02 &-& 20.54 & 22.67 &  21.88 & 22.65 && 24.47(4) \\ 
                        &  $\Delta E_{t,c}$ & 0.10  & -     &-& 0.05 &-& - & 0.02 & 0.09 & 0.04 && 0.15(4)  \\ \hline
  \multirow{6}{*}{m-HfO$_2$}  &  $a$  & 5.03 & 5.07 &5.39& 5.09&5.15 & 5.14 & 5.07 & 5.10 & 5.05 &5.12& 5.09(1)  \\
                              &  $b$  & 5.08 & 5.12 &5.33& 5.15 &5.19& 5.19 & 5.12 & 5.15 & 5.10 &5.16& 5.15(1)  \\
                              &  $c$ & 5.21 & 5.25 &5.49& 5.28 &5.33& 5.32 & 5.25 & 5.28 & 5.23 &5.28& 5.28(2)  \\
                              & $B_0$  & 275 & 253 &-& 248 &238& 223 & 270 & 255 & 269 &-&276(3)    \\ 
                         &  $E_{coh}$  & 27.07 & 24.99 &-& 23.15 &-& 20.63 & 22.82 & 22.02 & 22.81 &-& 24.64(3) \\ 
                         &  $\Delta E_{m,t}$ & 0.14 & - &-& 0.13 &-& - & 0.15 & 0.14 & 0.16 &-& 0.17(5) \\ \hline \hline
\end{tabular}
\begin{flushleft}
$^1$Ref.~\onlinecite{li13}.\\
$^2$Ref.~\onlinecite{beltran2008}.\\
$^3$Ref.~\onlinecite{lyons2011}.\\
\end{flushleft}
\end{table*}

\begin{table*}[t]
\centering
\caption{Results for structural parameters, bulk modulus, and cohesive energy for hafnia polymorphs from our DFT HSE06 calculations together with DMC data and experimental values.} 
\label{tab:HfO2_EOS2}
\begin{tabular}{ccccccccc}
\hline \hline
 phase &                                   & HSE06&  DMC    & Exp.$^1$ &Exp.$^2$ & Exp.$^3$ &   Exp.$^4$&Exp.$^5$ \\ \hline
  \multirow{2}{*}{c-HfO$_2$}  &  $a$ (\AA) & 5.00 & 5.04(1)      & 5.08 - 5.14     &-        & -        & - &-\\
                    &  $E_{coh}$ (eV/f.u.)  & 22.61& 24.32(3)       & -        &-        & -        &  - &-\\ \hline
  \multirow{3}{*}{t-HfO$_2$}  &  $a$        & 3.57  & 3.58(2)      & 3.63 - 3.66        &3.64     & -  & -&- \\
                              &  $c$        & 5.19  & 5.20(3)      & 5.25 - 5.33 &5.18     & -        & -&- \\
                          & $c$/$\sqrt{2}a$ & 1.03 & 1.03         &-         &1.01     & -        & -&-  \\ \hline
  \multirow{4}{*}{m-HfO$_2$}  &  $a$  & 5.05 & 5.09(1)             &-         &-       & 5.12    & - & 5.12   \\
                              &  $b$  & 5.10  & 5.15(1)            & -        & -       & 5.17    & -& 5.18\\
                              &  $c$ & 5.23 & 5.28(2)              & -        &-        & 5.30    & -& 5.29\\
                              & $B_0$ & 269 & 276(3)               & -        &-        & -       &  284(30)&- \\ \hline
\end{tabular}
\begin{flushleft}
$^1$Ref.~\onlinecite{wang92} at high temperatures.\\
$^2$Ref.~\onlinecite{adams91} powder diffraction at room T. The data for t-HfO$_2$ are quoted in Ref.~\onlinecite{terki08}.\\
$^3$Ref.~\onlinecite{ruh70} single crystal at room T.\\
$^4$Ref.~\onlinecite{desgreniers99} polycrystaline at room T.\\
$^5$Ref.~\onlinecite{hann85} powder diffraction at room T.
\end{flushleft}
\end{table*}

\begin{figure*}[t]
 \includegraphics[width=7 in]{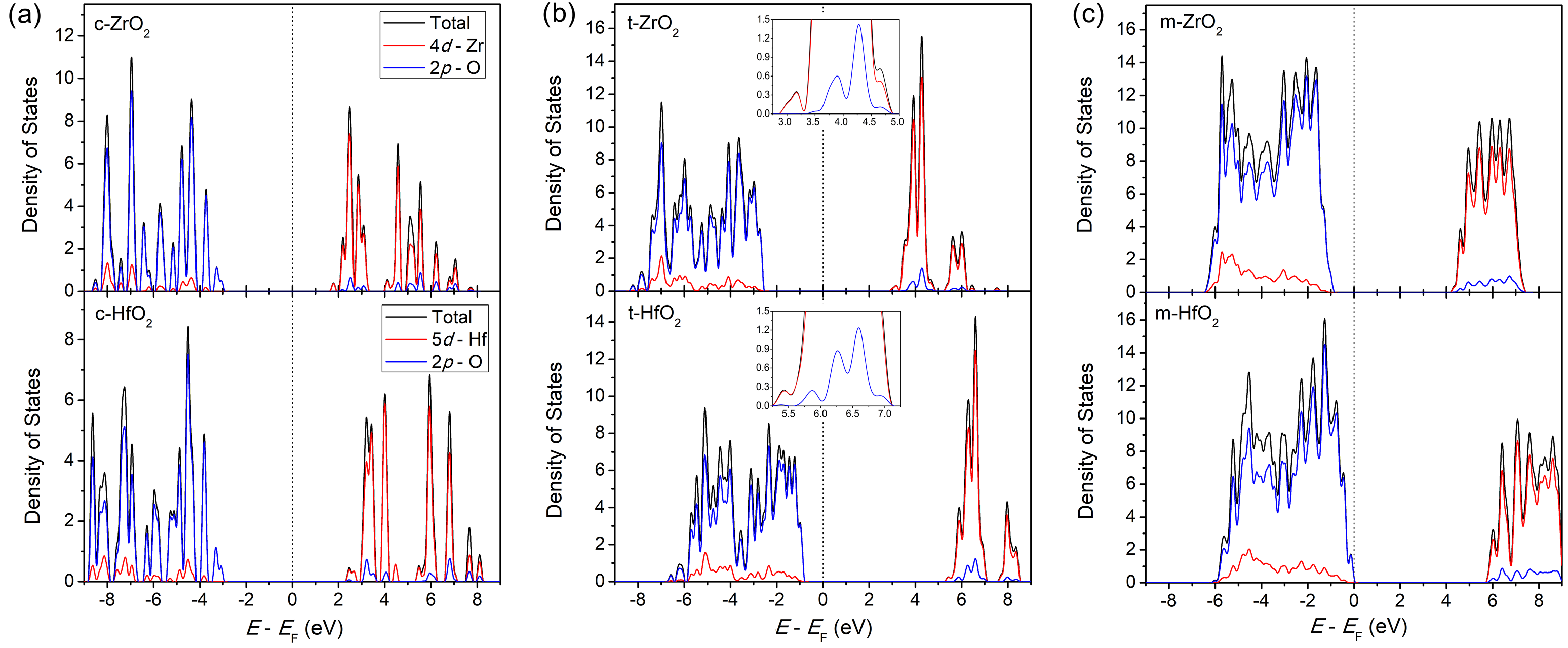}
 \caption{DFT density of states (DOS) and projected DOS for (a) cubic, (b) tetragonal, and (c) monoclinic phases of zirconia (top row) and hafnia (bottom row) obtained using the DFT HSE06 hybrid functional and QE. The insets in the center panels for the tetragonal phases show enlargements of DOS and PDOS near the conduction band edge.}
 \label{fig:HSE_DOS}
\end{figure*}

Figure~\ref{fig:EOS_HfO2} shows the computed DMC total energy as function of unit cell volume for the three HfO$_2$ polymorphs. The equilibrium lattice parameters, cohesive energy and bulk modulus estimated from Vinet fits are listed in Table~\ref{tab:HfO2_EOS} together with other representative DFT-based calculations from the literature. As expected LDA overbinds and its lattice parameters are consistently too small, while the lattice parameters from our GGA and the PBE calculation in Ref.~\cite{wang92} are in much better  agreement with the DMC values. LDA+U increases the lattice parameters but still overbinds - this is probably because the U-parameters from our optimization are quite small, as explained earlier. The lattice parameters from the hybrid functionals are in good agreement with the DMC values. The bulk moduli from the hybrid functionals are in quit good agreement with but consistently smaller than the DMC values, again with the poorest agreement for the cubic phase.  As was the case for zirconia, LDA yields bulk moduli in good agreement with DMC, while LDA+U reduces the overbinding but substantially softens the bulk modulus. It is interesting to note that for hafnia, the DMC bulk modulus for the tetragonal phase is smaller than for the monoclinic phase, although they are both equal to within one standard deviation; the DFT hybrid functionals roughly captures this as well. In contrast, the bulk modulus for tetragonal zirconia is significantly larger than the bulk modulus for monoclinic zirconia. This and the shallow minimum in energy vs. $d_z$ for t-HfO$_2$ compared to t-ZrO$_2$ indicate subtle differences in the energetics between the two compounds in spite of their superficial chemical equivalence. We speculate that this stems from the difference between the 5d Hf orbitals and the Zr 4d orbitals that lead to different hybridization with  oxygen 2p orbitals. Figure~\ref{fig:HSE_DOS} shows our DFT HSE06-calculated density of states (DOS) and projected densities of states (PDOS) for the zirconia and hafnia polymorphs. The PDOS for the tetragonal phases show a shift upward in energy of the oxygen 2p relative to the valence band edge in t-ZrO$_2$ compared to t-HfO$_2$, consistent with a difference in the hybridization (see insets in center panels of Fig.~\ref{fig:HSE_DOS}). All DFT calculations in Table~\ref{tab:HfO2_EOS} get the energetic ordering right, with the monoclinic phase having the lowest energy, followed by the tetragonal and cubic ones. The cohesive energies for the hybrid functionals are a bit too small and slightly worse compared to DMC than was the case for the zirconia polymorphs. The energy differences between the tetragonal and cubic phases are quite good for all DFT flavors, while they underestimate the energy difference between cubic and tetragonal phases. Again, the energy difference are quite small so small errors in the cohesive energies will give rise to relatively large errors in the energy differences.

\subsubsection{Hafnia - comparison with experiments}
\begin{figure}[t]
 \includegraphics[width=3.3 in]{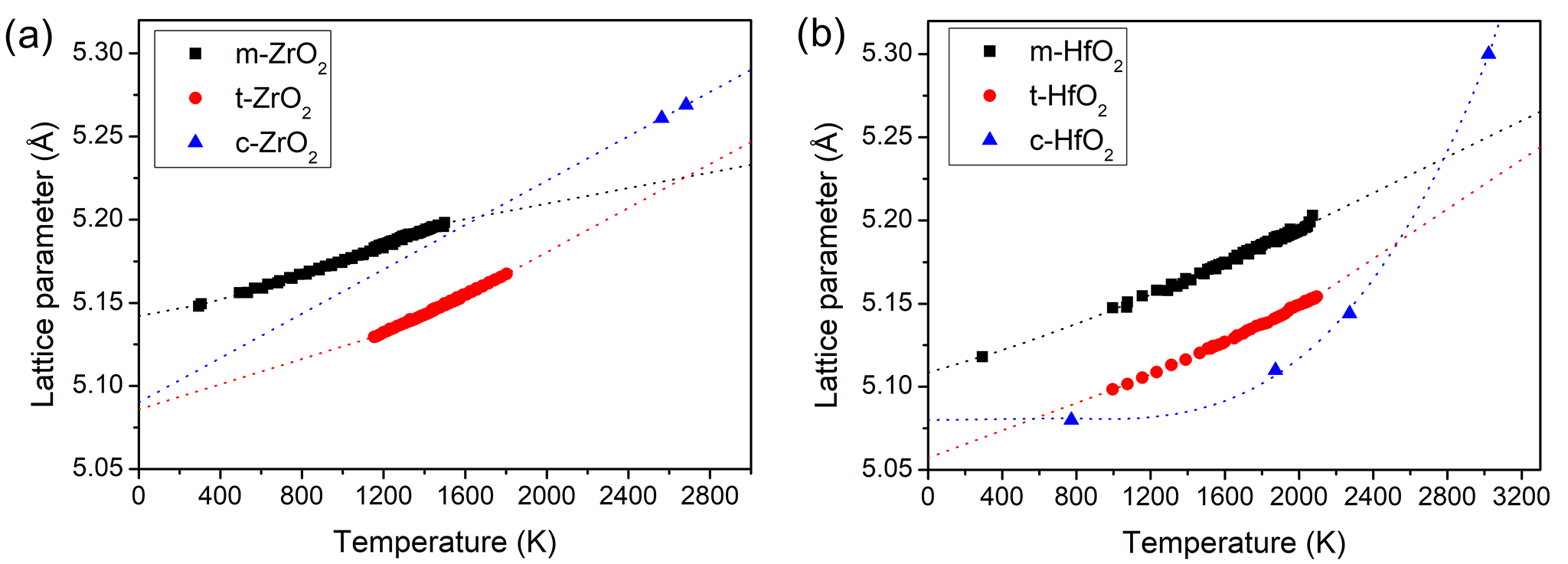}
 \caption{Experimentally measured lattice parameter $a$ as function of temperature for (a) zirconia and (b) hafnia polymorphs. We used cubic and polynomials to extrapolate lattice parameters to 0~K except for c-ZrO$_2$, for which we used a linear fit. The experimental lattice parameters are from Refs.~\onlinecite{haggerty14,shanshoury70,glushkova85,senft83,buckley68,aldebert85}. Note that for the tetragonal phases, we are reporting a value for a that is $\sqrt{2}$ larger than the values in Tables~IV - VII.}
 \label{fig:expansion}
\end{figure}

For the low-temperature monoclinic phase, the DMC result for the bulk modulus is in good agreement with the corresponding experimental value. However, the DFT HSE06 and DMC lattice parameters are smaller than the experimental ones because of thermal expansion at the experimental conditions. It is instructive to compare calculated and measured lattice parameters for hcp Hf and hafnia. 
Experimental lattice parameters for both hcp Hf and m-HfO$_2$ were measured at room temperature\cite{hann85,desgreniers99}. 
In contrast to the smaller DMC and DFT lattice parameters for m-HfO$_2$ (with the exception of GGA+U), the GGA and DMC lattice parameters for hcp Hf (see Table~\ref{tab:DMC_bulk}) are in very good agreement with the experimental value.
Hcp Hf is known\cite{perry11} to have a small thermal expansion coefficient, about $5.9\times10^{-6}$~K$^{-1}$. In contrast, the thermal expansion coefficient for m-HfO$_2$ has been observed\cite{haggerty14} to vary significantly with temperature, and is considerably larger than for hcp Hf, as much as $32\times10^{-6}$~K$^{-1}$.  
This relatively high sensitivity of the thermal expansion coefficient of m-HfO$_2$, and therefore of its volume, on temperature supports our argument that the underestimated  DMC and DFT lattice parameters are due to volume expansion of m-HfO$_2$ in the experimental measurements performed at room temperature.
As a result of this thermal expansion in hafnia at high temperatures, it is also not surprising to observe larger experimental lattice parameters of the tetragonal and cubic phases measured at 1,600~K (t-HfO$_2$)\cite{curtis54,haggerty14} and 800~K (c-HfO$_2$)\cite{shanshoury70}, respectively, than our DFT or DMC results, which represent properties of  a (non-physical or metastable) zero-temperature phase. 
This is particularly the case for X-ray diffraction measurements of t-HfO$_2$ which were performed at 1,600~K, and which shows the largest discrepancy, about 1.6~\%, between the DMC and experimental values among three polymorphs. We attempted to correct for the thermal expansion in the experimental measurements by extrapolating the measured lattice constants to 0~K [see Fig.~\ref{fig:expansion}) (b)]. For m-HfO$_2$, we obtain an extrapolated value of $a$=5.11~{\AA} and $c$=5.28~{\AA} for cubic fits, and for t-HfO$_2$ we obtain $a$=3.58~{\AA}, $c$=5.17~{\AA} (5.07~{\AA}) for quadratic (cubic) fits, in good agreements with the DMC values. The extrapolation for the lattice parameter of c-HfO$_2$ is considerably more uncertain because of the sparsity in temperature-dependent measurements of the cubic lattice parameter. We obtain an extrapolated value of about 5.08~{\AA}, which is in the lower range of the measurements at high temperatures.


\section{\label{sec:summary}Summary and Conclusions}
We have evaluated bulk properties (lattice constants, bulk modulus, and cohesive energy) of the monoclinic, tetragonal, and cubic phases of ZrO$_2$ and of HfO$_2$ using diffusion Monte Carlo methods, and compared the diffusion Monte Carlo results with those of ours as well as in the literature obtained using various DFT schemes (LDA, GGA, LDA+U, GGA+U, hybrid functionals PBE0, B3LYP, and HSE06) . We also examined the distortion of oxygen columns in the tetragonal phases of ZrO$_2$ and HfO$_2$ at the experimental unit cell volume. The obtained DMC values for lattice parameters and distortion of the oxygen columns are in good qualitative and quantitative agreement with experimental values extrapolated to 0~K. Such extrapolations are more reliable for the zirconia phases than for the hafnia ones, as there exist more data for zirconia, both measurements of lattice parameters for a range of temperatures as well as low- or room-temperature measurements of nanocrystalline t-ZrO$_2$; such data are considerably sparser for the hafnia phases. The DFT results, in comparison with the DMC results, are inconsistent in that no DFT scheme generally agrees with DMC both for lattice parameters and bulk modulus. For example, the lattice constants from the hybrid functionals PBE0 and HSE06 agree well with the DMC ones, while the bulk moduli are smaller than the DMC ones; the LDA lattice parameters are too small while the bulk moduli are in rather good agreement with the DMC ones. However, the hybrid functionals PBE0 and HSE06 in general yield results in reasonable agreement with DMC. We note that such agreements should not be taken for granted: even HSE06 has been shown to give dramatically incorrect results for correlated transition metal oxides\cite{grau2012}. The calculated bulk moduli, both DFT and DMC, are larger than experimentally measured ones, especially for the zirconia polymorphs, although we note that there is a lack of experimental data for the bulk modulus of tetragonal and cubic hafnia. 
Direct comparison between calculated bulk moduli that do not include finite-temperature effects and experiments is difficult and perhaps not very meaningful; rather, we hope that the DMC calculations of bulk modulus will be useful to benchmark DFT calculations. There are many examples of DFT calculations of bulk moduli for zirconia and hafnia that do not include finite-temperature corrections\cite{terki2006,milman2009,terki08}. These results have large spreads depending on what code and what flavor of DFT was used. Therefore, we believe there is value in our DMC values to serve as a guide for future DFT studies, both for bulk modulus as well as for cohesive energy (and lattice parameters). 

The validation of our pseudopotentials for atomic and bulk hcp Zr and Hf shows that the pseudopotentials give excellent results, and errors arising from the DMC pseudopotentials can be neglected. A potential source of error in the DMC calculation is that the nodal surface may not be optimal. We routinely use LDA+U and PBE+U to minimize the DMC ground state energy as function of U in the trial wavefunction. LDA+U tends consistently to give lower DMC energy than PBE+U, which is why we used LDA+U here (we did not here present the results for the PBE+U optimization of U) to obtain trial wavefunctions for the QMC calculations. This is also consistent with other DMC work on transition metal oxides\cite{BenaliTi4O7,shin17,shin17NiO} where such optimization of the nodal surface has shown to be a very efficient way in general yielding results of structural parameters in very good agreement with available experimental data. While this optimization does not guarantee that the true ground state has been obtained, our experience has shown that it works very well. Future methods under development will include expansion of the trial wavefunction in a large number of determinants selected perturbatively. It is expected that those methods, when implemented, will obtain extremely good nodal surfaces.\\

\begin{acknowledgements}
An award of computer time was provided by the Innovative and Novel Computational Impact on Theory and Experiment (INCITE) program. This research has used resources of the Argonne Leadership Computing Facility, which is a DOE Office of Science User Facility supported under Contract DE-AC02-06CH11357.  AB, YL, HS and OH were supported by the U.S. Department of Energy, Office of Science, Basic Energy Sciences, Materials Sciences and Engineering Division, as part of the Computational Materials Sciences Program and Center for Predictive Simulation of Functional Materials. EC acknowledges support from a DOE Computational Science Graduate Fellowship. We gratefully acknowledge the computing resources provided on Blues and Bebop, high-performance computing clusters operated by the Laboratory Computing Resource Center at Argonne National Laboratory.
\end{acknowledgements}
\bibliography{all}

\begin{thebibliography}{96}%
\makeatletter
\providecommand \@ifxundefined [1]{%
 \@ifx{#1\undefined}
}%
\providecommand \@ifnum [1]{%
 \ifnum #1\expandafter \@firstoftwo
 \else \expandafter \@secondoftwo
 \fi
}%
\providecommand \@ifx [1]{%
 \ifx #1\expandafter \@firstoftwo
 \else \expandafter \@secondoftwo
 \fi
}%
\providecommand \natexlab [1]{#1}%
\providecommand \enquote  [1]{``#1''}%
\providecommand \bibnamefont  [1]{#1}%
\providecommand \bibfnamefont [1]{#1}%
\providecommand \citenamefont [1]{#1}%
\providecommand \href@noop [0]{\@secondoftwo}%
\providecommand \href [0]{\begingroup \@sanitize@url \@href}%
\providecommand \@href[1]{\@@startlink{#1}\@@href}%
\providecommand \@@href[1]{\endgroup#1\@@endlink}%
\providecommand \@sanitize@url [0]{\catcode `\\12\catcode `\$12\catcode
  `\&12\catcode `\#12\catcode `\^12\catcode `\_12\catcode `\%12\relax}%
\providecommand \@@startlink[1]{}%
\providecommand \@@endlink[0]{}%
\providecommand \url  [0]{\begingroup\@sanitize@url \@url }%
\providecommand \@url [1]{\endgroup\@href {#1}{\urlprefix }}%
\providecommand \urlprefix  [0]{URL }%
\providecommand \Eprint [0]{\href }%
\providecommand \doibase [0]{http://dx.doi.org/}%
\providecommand \selectlanguage [0]{\@gobble}%
\providecommand \bibinfo  [0]{\@secondoftwo}%
\providecommand \bibfield  [0]{\@secondoftwo}%
\providecommand \translation [1]{[#1]}%
\providecommand \BibitemOpen [0]{}%
\providecommand \bibitemStop [0]{}%
\providecommand \bibitemNoStop [0]{.\EOS\space}%
\providecommand \EOS [0]{\spacefactor3000\relax}%
\providecommand \BibitemShut  [1]{\csname bibitem#1\endcsname}%
\let\auto@bib@innerbib\@empty
\bibitem [{\citenamefont {Manicone}\ \emph {et~al.}(2007)\citenamefont
  {Manicone}, \citenamefont {Iommetti},\ and\ \citenamefont
  {Raffaelli}}]{manicone2007}%
  \BibitemOpen
  \bibfield  {author} {\bibinfo {author} {\bibfnamefont {P.~F.}\ \bibnamefont
  {Manicone}}, \bibinfo {author} {\bibfnamefont {P.~R.}\ \bibnamefont
  {Iommetti}}, \ and\ \bibinfo {author} {\bibfnamefont {L.}~\bibnamefont
  {Raffaelli}},\ }\href@noop {} {\bibfield  {journal} {\bibinfo  {journal}
  {Journal of dentistry}\ }\textbf {\bibinfo {volume} {35}},\ \bibinfo {pages}
  {819} (\bibinfo {year} {2007})}\BibitemShut {NoStop}%
\bibitem [{\citenamefont {Bocanegra-Bernal}\ and\ \citenamefont
  {De~La~Torre}(2002)}]{bocanegra2002}%
  \BibitemOpen
  \bibfield  {author} {\bibinfo {author} {\bibfnamefont {M.}~\bibnamefont
  {Bocanegra-Bernal}}\ and\ \bibinfo {author} {\bibfnamefont {S.~D.}\
  \bibnamefont {De~La~Torre}},\ }\href@noop {} {\bibfield  {journal} {\bibinfo
  {journal} {Journal of materials science}\ }\textbf {\bibinfo {volume} {37}},\
  \bibinfo {pages} {4947} (\bibinfo {year} {2002})}\BibitemShut {NoStop}%
\bibitem [{\citenamefont {Motta}\ \emph {et~al.}(2015)\citenamefont {Motta},
  \citenamefont {Couet},\ and\ \citenamefont {Comstock}}]{motta2015}%
  \BibitemOpen
  \bibfield  {author} {\bibinfo {author} {\bibfnamefont {A.~T.}\ \bibnamefont
  {Motta}}, \bibinfo {author} {\bibfnamefont {A.}~\bibnamefont {Couet}}, \ and\
  \bibinfo {author} {\bibfnamefont {R.~J.}\ \bibnamefont {Comstock}},\
  }\href@noop {} {\bibfield  {journal} {\bibinfo  {journal} {Annual Review of
  Materials Research}\ }\textbf {\bibinfo {volume} {45}},\ \bibinfo {pages}
  {311} (\bibinfo {year} {2015})}\BibitemShut {NoStop}%
\bibitem [{\citenamefont {Saxena}\ and\ \citenamefont
  {Mittal}(1975)}]{saxena75}%
  \BibitemOpen
  \bibfield  {author} {\bibinfo {author} {\bibfnamefont {A.~N.}\ \bibnamefont
  {Saxena}}\ and\ \bibinfo {author} {\bibfnamefont {K.~L.}\ \bibnamefont
  {Mittal}},\ }\href@noop {} {\bibfield  {journal} {\bibinfo  {journal} {J.
  Appl. Phys.}\ }\textbf {\bibinfo {volume} {46}},\ \bibinfo {pages} {2788}
  (\bibinfo {year} {1975})}\BibitemShut {NoStop}%
\bibitem [{\citenamefont {Edlou}\ \emph {et~al.}(1993)\citenamefont {Edlou},
  \citenamefont {Smajkiewicz},\ and\ \citenamefont {Al-Jumaily}}]{edlou93}%
  \BibitemOpen
  \bibfield  {author} {\bibinfo {author} {\bibfnamefont {S.~M.}\ \bibnamefont
  {Edlou}}, \bibinfo {author} {\bibfnamefont {A.}~\bibnamefont {Smajkiewicz}},
  \ and\ \bibinfo {author} {\bibfnamefont {G.~A.}\ \bibnamefont {Al-Jumaily}},\
  }\href@noop {} {\bibfield  {journal} {\bibinfo  {journal} {Appl. Opt.}\
  }\textbf {\bibinfo {volume} {32}},\ \bibinfo {pages} {5601} (\bibinfo {year}
  {1993})}\BibitemShut {NoStop}%
\bibitem [{\citenamefont {Lee}\ \emph {et~al.}(2000)\citenamefont {Lee},
  \citenamefont {Kang}, \citenamefont {Nieh}, \citenamefont {Qi},\ and\
  \citenamefont {Lee}}]{lee00}%
  \BibitemOpen
  \bibfield  {author} {\bibinfo {author} {\bibfnamefont {B.~H.}\ \bibnamefont
  {Lee}}, \bibinfo {author} {\bibfnamefont {L.}~\bibnamefont {Kang}}, \bibinfo
  {author} {\bibfnamefont {R.}~\bibnamefont {Nieh}}, \bibinfo {author}
  {\bibfnamefont {W.-J.}\ \bibnamefont {Qi}}, \ and\ \bibinfo {author}
  {\bibfnamefont {J.~C.}\ \bibnamefont {Lee}},\ }\href@noop {} {\bibfield
  {journal} {\bibinfo  {journal} {Appl. Phys. Lett.}\ }\textbf {\bibinfo
  {volume} {76}},\ \bibinfo {pages} {1926} (\bibinfo {year}
  {2000})}\BibitemShut {NoStop}%
\bibitem [{\citenamefont {Kang}\ \emph {et~al.}(2000)\citenamefont {Kang},
  \citenamefont {Lee}, \citenamefont {Qi}, \citenamefont {Jeon}, \citenamefont
  {Nieh}, \citenamefont {Gopalan}, \citenamefont {Onishi},\ and\ \citenamefont
  {Lee}}]{kang00}%
  \BibitemOpen
  \bibfield  {author} {\bibinfo {author} {\bibfnamefont {L.}~\bibnamefont
  {Kang}}, \bibinfo {author} {\bibfnamefont {B.~H.}\ \bibnamefont {Lee}},
  \bibinfo {author} {\bibfnamefont {W.-J.}\ \bibnamefont {Qi}}, \bibinfo
  {author} {\bibfnamefont {Y.}~\bibnamefont {Jeon}}, \bibinfo {author}
  {\bibfnamefont {R.}~\bibnamefont {Nieh}}, \bibinfo {author} {\bibfnamefont
  {S.}~\bibnamefont {Gopalan}}, \bibinfo {author} {\bibfnamefont
  {K.}~\bibnamefont {Onishi}}, \ and\ \bibinfo {author} {\bibfnamefont {J.~C.}\
  \bibnamefont {Lee}},\ }\href@noop {} {\bibfield  {journal} {\bibinfo
  {journal} {IEEE Electron Device Lett.}\ }\textbf {\bibinfo {volume} {21}},\
  \bibinfo {pages} {181} (\bibinfo {year} {2000})}\BibitemShut {NoStop}%
\bibitem [{\citenamefont {Al-Kuhaili}(2004)}]{kuhaili04}%
  \BibitemOpen
  \bibfield  {author} {\bibinfo {author} {\bibfnamefont {M.~F.}\ \bibnamefont
  {Al-Kuhaili}},\ }\href@noop {} {\bibfield  {journal} {\bibinfo  {journal}
  {Opt. Mater.}\ }\textbf {\bibinfo {volume} {27}},\ \bibinfo {pages} {383}
  (\bibinfo {year} {2004})}\BibitemShut {NoStop}%
\bibitem [{\citenamefont {Robertson}(2006)}]{robertson06}%
  \BibitemOpen
  \bibfield  {author} {\bibinfo {author} {\bibfnamefont {J.}~\bibnamefont
  {Robertson}},\ }\href@noop {} {\bibfield  {journal} {\bibinfo  {journal}
  {Rep. Prog. Phys.}\ }\textbf {\bibinfo {volume} {69}},\ \bibinfo {pages}
  {327} (\bibinfo {year} {2006})}\BibitemShut {NoStop}%
\bibitem [{\citenamefont {Khoshman}\ \emph {et~al.}(2008)\citenamefont
  {Khoshman}, \citenamefont {Khan},\ and\ \citenamefont
  {Kordesch}}]{khoshman08}%
  \BibitemOpen
  \bibfield  {author} {\bibinfo {author} {\bibfnamefont {J.~M.}\ \bibnamefont
  {Khoshman}}, \bibinfo {author} {\bibfnamefont {A.}~\bibnamefont {Khan}}, \
  and\ \bibinfo {author} {\bibfnamefont {M.~E.}\ \bibnamefont {Kordesch}},\
  }\href@noop {} {\bibfield  {journal} {\bibinfo  {journal} {Surf. Coat.
  Technol.}\ }\textbf {\bibinfo {volume} {202}},\ \bibinfo {pages} {2500}
  (\bibinfo {year} {2008})}\BibitemShut {NoStop}%
\bibitem [{\citenamefont {Kittl}\ \emph {et~al.}(2009)\citenamefont {Kittl},
  \citenamefont {Opsomer}, \citenamefont {Popovici}, \citenamefont {Menou},
  \citenamefont {Kaczer}, \citenamefont {Wang}, \citenamefont {Adelmann},
  \citenamefont {Pawlak}, \citenamefont {Tomida}, \citenamefont {Rothschild},
  \citenamefont {Govoreanu}, \citenamefont {Degraeve}, \citenamefont
  {Schaekers}, \citenamefont {Zahid}, \citenamefont {Delabie}, \citenamefont
  {Meersschaut}, \citenamefont {Polspoel}, \citenamefont {Clima}, \citenamefont
  {Pourtois}, \citenamefont {Knaepen}, \citenamefont {Detavernier},
  \citenamefont {Afanas'ev}, \citenamefont {Blomberg}, \citenamefont
  {Pierreux}, \citenamefont {Swerts}, \citenamefont {Fischer}, \citenamefont
  {Maes}, \citenamefont {Manger}, \citenamefont {Vandervorst}, \citenamefont
  {Conrad}, \citenamefont {Franquet}, \citenamefont {Favia}, \citenamefont
  {Bender}, \citenamefont {Brijs}, \citenamefont {Elshocht}, \citenamefont
  {Jurczak}, \citenamefont {Houdt},\ and\ \citenamefont {Wouters}}]{kittl09}%
  \BibitemOpen
  \bibfield  {author} {\bibinfo {author} {\bibfnamefont {J.~A.}\ \bibnamefont
  {Kittl}}, \bibinfo {author} {\bibfnamefont {K.}~\bibnamefont {Opsomer}},
  \bibinfo {author} {\bibfnamefont {M.}~\bibnamefont {Popovici}}, \bibinfo
  {author} {\bibfnamefont {N.}~\bibnamefont {Menou}}, \bibinfo {author}
  {\bibfnamefont {B.}~\bibnamefont {Kaczer}}, \bibinfo {author} {\bibfnamefont
  {X.~P.}\ \bibnamefont {Wang}}, \bibinfo {author} {\bibfnamefont
  {C.}~\bibnamefont {Adelmann}}, \bibinfo {author} {\bibfnamefont {M.~A.}\
  \bibnamefont {Pawlak}}, \bibinfo {author} {\bibfnamefont {K.}~\bibnamefont
  {Tomida}}, \bibinfo {author} {\bibfnamefont {A.}~\bibnamefont {Rothschild}},
  \bibinfo {author} {\bibfnamefont {B.}~\bibnamefont {Govoreanu}}, \bibinfo
  {author} {\bibfnamefont {R.}~\bibnamefont {Degraeve}}, \bibinfo {author}
  {\bibfnamefont {M.}~\bibnamefont {Schaekers}}, \bibinfo {author}
  {\bibfnamefont {M.}~\bibnamefont {Zahid}}, \bibinfo {author} {\bibfnamefont
  {A.}~\bibnamefont {Delabie}}, \bibinfo {author} {\bibfnamefont
  {J.}~\bibnamefont {Meersschaut}}, \bibinfo {author} {\bibfnamefont
  {W.}~\bibnamefont {Polspoel}}, \bibinfo {author} {\bibfnamefont
  {S.}~\bibnamefont {Clima}}, \bibinfo {author} {\bibfnamefont
  {G.}~\bibnamefont {Pourtois}}, \bibinfo {author} {\bibfnamefont
  {W.}~\bibnamefont {Knaepen}}, \bibinfo {author} {\bibfnamefont
  {C.}~\bibnamefont {Detavernier}}, \bibinfo {author} {\bibfnamefont {V.~V.}\
  \bibnamefont {Afanas'ev}}, \bibinfo {author} {\bibfnamefont {T.}~\bibnamefont
  {Blomberg}}, \bibinfo {author} {\bibfnamefont {D.}~\bibnamefont {Pierreux}},
  \bibinfo {author} {\bibfnamefont {J.}~\bibnamefont {Swerts}}, \bibinfo
  {author} {\bibfnamefont {P.}~\bibnamefont {Fischer}}, \bibinfo {author}
  {\bibfnamefont {J.~W.}\ \bibnamefont {Maes}}, \bibinfo {author}
  {\bibfnamefont {D.}~\bibnamefont {Manger}}, \bibinfo {author} {\bibfnamefont
  {W.}~\bibnamefont {Vandervorst}}, \bibinfo {author} {\bibfnamefont
  {T.}~\bibnamefont {Conrad}}, \bibinfo {author} {\bibfnamefont
  {A.}~\bibnamefont {Franquet}}, \bibinfo {author} {\bibfnamefont
  {P.}~\bibnamefont {Favia}}, \bibinfo {author} {\bibfnamefont
  {H.}~\bibnamefont {Bender}}, \bibinfo {author} {\bibfnamefont
  {B.}~\bibnamefont {Brijs}}, \bibinfo {author} {\bibfnamefont {S.~V.}\
  \bibnamefont {Elshocht}}, \bibinfo {author} {\bibfnamefont {M.}~\bibnamefont
  {Jurczak}}, \bibinfo {author} {\bibfnamefont {J.~V.}\ \bibnamefont {Houdt}},
  \ and\ \bibinfo {author} {\bibfnamefont {D.~J.}\ \bibnamefont {Wouters}},\
  }\href@noop {} {\bibfield  {journal} {\bibinfo  {journal} {Microelectron.
  Eng.}\ }\textbf {\bibinfo {volume} {86}},\ \bibinfo {pages} {1789} (\bibinfo
  {year} {2009})}\BibitemShut {NoStop}%
\bibitem [{\citenamefont {Choi}\ \emph {et~al.}(2011)\citenamefont {Choi},
  \citenamefont {Mao},\ and\ \citenamefont {Chang}}]{choi11}%
  \BibitemOpen
  \bibfield  {author} {\bibinfo {author} {\bibfnamefont {J.~H.}\ \bibnamefont
  {Choi}}, \bibinfo {author} {\bibfnamefont {Y.}~\bibnamefont {Mao}}, \ and\
  \bibinfo {author} {\bibfnamefont {J.~P.}\ \bibnamefont {Chang}},\ }\href@noop
  {} {\bibfield  {journal} {\bibinfo  {journal} {Mater. Sci. Eng. R}\ }\textbf
  {\bibinfo {volume} {72}},\ \bibinfo {pages} {97} (\bibinfo {year}
  {2011})}\BibitemShut {NoStop}%
\bibitem [{\citenamefont {Wong}\ \emph {et~al.}(2012)\citenamefont {Wong},
  \citenamefont {Lee}, \citenamefont {Yu}, \citenamefont {Chen}, \citenamefont
  {Wu}, \citenamefont {Chen}, \citenamefont {Lee}, \citenamefont {Chen},\ and\
  \citenamefont {Tsai}}]{wong12}%
  \BibitemOpen
  \bibfield  {author} {\bibinfo {author} {\bibfnamefont {H.-S.~P.}\
  \bibnamefont {Wong}}, \bibinfo {author} {\bibfnamefont {H.-Y.}\ \bibnamefont
  {Lee}}, \bibinfo {author} {\bibfnamefont {S.}~\bibnamefont {Yu}}, \bibinfo
  {author} {\bibfnamefont {Y.-S.}\ \bibnamefont {Chen}}, \bibinfo {author}
  {\bibfnamefont {Y.}~\bibnamefont {Wu}}, \bibinfo {author} {\bibfnamefont
  {P.-S.}\ \bibnamefont {Chen}}, \bibinfo {author} {\bibfnamefont
  {B.}~\bibnamefont {Lee}}, \bibinfo {author} {\bibfnamefont {F.~T.}\
  \bibnamefont {Chen}}, \ and\ \bibinfo {author} {\bibfnamefont {M.-J.}\
  \bibnamefont {Tsai}},\ }\href@noop {} {\bibfield  {journal} {\bibinfo
  {journal} {Proc. IEEE}\ }\textbf {\bibinfo {volume} {100}},\ \bibinfo {pages}
  {1951} (\bibinfo {year} {2012})}\BibitemShut {NoStop}%
\bibitem [{\citenamefont {Chertihin}\ and\ \citenamefont
  {Andrews}(1995)}]{chertihin95}%
  \BibitemOpen
  \bibfield  {author} {\bibinfo {author} {\bibfnamefont {G.~V.}\ \bibnamefont
  {Chertihin}}\ and\ \bibinfo {author} {\bibfnamefont {L.}~\bibnamefont
  {Andrews}},\ }\href@noop {} {\bibfield  {journal} {\bibinfo  {journal} {J.
  Phys. Chem.}\ }\textbf {\bibinfo {volume} {99}},\ \bibinfo {pages} {6356}
  (\bibinfo {year} {1995})}\BibitemShut {NoStop}%
\bibitem [{\citenamefont {Brugh}\ \emph {et~al.}(1999)\citenamefont {Brugh},
  \citenamefont {Suenram},\ and\ \citenamefont {Stevens}}]{brugh99}%
  \BibitemOpen
  \bibfield  {author} {\bibinfo {author} {\bibfnamefont {D.~J.}\ \bibnamefont
  {Brugh}}, \bibinfo {author} {\bibfnamefont {R.~D.}\ \bibnamefont {Suenram}},
  \ and\ \bibinfo {author} {\bibfnamefont {W.~J.}\ \bibnamefont {Stevens}},\
  }\href@noop {} {\bibfield  {journal} {\bibinfo  {journal} {J. Chem. Phys.}\
  }\textbf {\bibinfo {volume} {111}},\ \bibinfo {pages} {3526} (\bibinfo {year}
  {1999})}\BibitemShut {NoStop}%
\bibitem [{\citenamefont {Lesarri}\ \emph {et~al.}(2012)\citenamefont
  {Lesarri}, \citenamefont {Suenram},\ and\ \citenamefont {Brugh}}]{lesarri12}%
  \BibitemOpen
  \bibfield  {author} {\bibinfo {author} {\bibfnamefont {A.}~\bibnamefont
  {Lesarri}}, \bibinfo {author} {\bibfnamefont {R.~D.}\ \bibnamefont
  {Suenram}}, \ and\ \bibinfo {author} {\bibfnamefont {D.}~\bibnamefont
  {Brugh}},\ }\href@noop {} {\bibfield  {journal} {\bibinfo  {journal} {J.
  Chem. Phys.}\ }\textbf {\bibinfo {volume} {117}},\ \bibinfo {pages} {9651}
  (\bibinfo {year} {2012})}\BibitemShut {NoStop}%
\bibitem [{\citenamefont {Kisi}\ and\ \citenamefont {Howard}(1998)}]{kisi98}%
  \BibitemOpen
  \bibfield  {author} {\bibinfo {author} {\bibfnamefont {E.~H.}\ \bibnamefont
  {Kisi}}\ and\ \bibinfo {author} {\bibfnamefont {C.~J.}\ \bibnamefont
  {Howard}},\ }\href@noop {} {\bibfield  {journal} {\bibinfo  {journal} {Key.
  Eng. Mater.}\ }\textbf {\bibinfo {volume} {153-154}},\ \bibinfo {pages} {1}
  (\bibinfo {year} {1998})}\BibitemShut {NoStop}%
\bibitem [{\citenamefont {Massalski}\ \emph {et~al.}(1990)\citenamefont
  {Massalski}, \citenamefont {Okamoto}, \citenamefont {Subramanian},\ and\
  \citenamefont {Kacprzak}}]{massalski90}%
  \BibitemOpen
  \bibfield  {author} {\bibinfo {author} {\bibfnamefont {T.~B.}\ \bibnamefont
  {Massalski}}, \bibinfo {author} {\bibfnamefont {H.}~\bibnamefont {Okamoto}},
  \bibinfo {author} {\bibfnamefont {P.~R.}\ \bibnamefont {Subramanian}}, \ and\
  \bibinfo {author} {\bibfnamefont {L.}~\bibnamefont {Kacprzak}},\ }\href@noop
  {} {\emph {\bibinfo {title} {Binary Alloy Phase Diagrams, 2nd Ed.}}}\
  (\bibinfo  {publisher} {ASM International},\ \bibinfo {year}
  {1990})\BibitemShut {NoStop}%
\bibitem [{\citenamefont {Baun}(1963)}]{baun63}%
  \BibitemOpen
  \bibfield  {author} {\bibinfo {author} {\bibfnamefont {W.~L.}\ \bibnamefont
  {Baun}},\ }\href@noop {} {\bibfield  {journal} {\bibinfo  {journal}
  {Science}\ }\textbf {\bibinfo {volume} {140}},\ \bibinfo {pages} {1330}
  (\bibinfo {year} {1963})}\BibitemShut {NoStop}%
\bibitem [{\citenamefont {Wolten}(1963)}]{wolten63}%
  \BibitemOpen
  \bibfield  {author} {\bibinfo {author} {\bibfnamefont {G.~M.}\ \bibnamefont
  {Wolten}},\ }\href@noop {} {\bibfield  {journal} {\bibinfo  {journal} {J. Am.
  Ceram. Soc.}\ }\textbf {\bibinfo {volume} {46}},\ \bibinfo {pages} {418}
  (\bibinfo {year} {1963})}\BibitemShut {NoStop}%
\bibitem [{\citenamefont {Gutowski}\ \emph {et~al.}(2002)\citenamefont
  {Gutowski}, \citenamefont {Jaffe}, \citenamefont {Liu}, \citenamefont
  {Stoker}, \citenamefont {Hegde}, \citenamefont {Rai},\ and\ \citenamefont
  {Tobin}}]{gutowski02}%
  \BibitemOpen
  \bibfield  {author} {\bibinfo {author} {\bibfnamefont {M.}~\bibnamefont
  {Gutowski}}, \bibinfo {author} {\bibfnamefont {J.~E.}\ \bibnamefont {Jaffe}},
  \bibinfo {author} {\bibfnamefont {C.-L.}\ \bibnamefont {Liu}}, \bibinfo
  {author} {\bibfnamefont {M.}~\bibnamefont {Stoker}}, \bibinfo {author}
  {\bibfnamefont {R.~I.}\ \bibnamefont {Hegde}}, \bibinfo {author}
  {\bibfnamefont {R.~S.}\ \bibnamefont {Rai}}, \ and\ \bibinfo {author}
  {\bibfnamefont {P.~J.}\ \bibnamefont {Tobin}},\ }\href@noop {} {\bibfield
  {journal} {\bibinfo  {journal} {Appl. Phys. Lett.}\ }\textbf {\bibinfo
  {volume} {80}},\ \bibinfo {pages} {1897} (\bibinfo {year}
  {2002})}\BibitemShut {NoStop}%
\bibitem [{\citenamefont {Ackermann}\ \emph {et~al.}(1975)\citenamefont
  {Ackermann}, \citenamefont {Rauh},\ and\ \citenamefont
  {Alexander}}]{ackermann1975}%
  \BibitemOpen
  \bibfield  {author} {\bibinfo {author} {\bibfnamefont {R.}~\bibnamefont
  {Ackermann}}, \bibinfo {author} {\bibfnamefont {E.}~\bibnamefont {Rauh}}, \
  and\ \bibinfo {author} {\bibfnamefont {C.}~\bibnamefont {Alexander}},\
  }\href@noop {} {\bibfield  {journal} {\bibinfo  {journal} {High Temperature
  Science}\ }\textbf {\bibinfo {volume} {7}},\ \bibinfo {pages} {304} (\bibinfo
  {year} {1975})}\BibitemShut {NoStop}%
\bibitem [{\citenamefont {Aldebert}\ and\ \citenamefont
  {Traverse}(1985)}]{aldebert85}%
  \BibitemOpen
  \bibfield  {author} {\bibinfo {author} {\bibfnamefont {P.}~\bibnamefont
  {Aldebert}}\ and\ \bibinfo {author} {\bibfnamefont {J.-P.}\ \bibnamefont
  {Traverse}},\ }\href@noop {} {\bibfield  {journal} {\bibinfo  {journal} {J.
  Am. Ceram. Soc.}\ }\textbf {\bibinfo {volume} {68}},\ \bibinfo {pages} {34}
  (\bibinfo {year} {1985})}\BibitemShut {NoStop}%
\bibitem [{\citenamefont {Howard}\ \emph {et~al.}(1988)\citenamefont {Howard},
  \citenamefont {Hill},\ and\ \citenamefont {Reichert}}]{howard88}%
  \BibitemOpen
  \bibfield  {author} {\bibinfo {author} {\bibfnamefont {C.~J.}\ \bibnamefont
  {Howard}}, \bibinfo {author} {\bibfnamefont {R.~J.}\ \bibnamefont {Hill}}, \
  and\ \bibinfo {author} {\bibfnamefont {B.~E.}\ \bibnamefont {Reichert}},\
  }\href@noop {} {\bibfield  {journal} {\bibinfo  {journal} {Acta Cryst. B}\
  }\textbf {\bibinfo {volume} {44}},\ \bibinfo {pages} {116} (\bibinfo {year}
  {1988})}\BibitemShut {NoStop}%
\bibitem [{\citenamefont {Stefanovich}\ \emph {et~al.}(1994)\citenamefont
  {Stefanovich}, \citenamefont {Shluger},\ and\ \citenamefont
  {Catlow}}]{stefanovich1994}%
  \BibitemOpen
  \bibfield  {author} {\bibinfo {author} {\bibfnamefont {E.}~\bibnamefont
  {Stefanovich}}, \bibinfo {author} {\bibfnamefont {A.~L.}\ \bibnamefont
  {Shluger}}, \ and\ \bibinfo {author} {\bibfnamefont {C.}~\bibnamefont
  {Catlow}},\ }\href@noop {} {\bibfield  {journal} {\bibinfo  {journal}
  {Physical Review B}\ }\textbf {\bibinfo {volume} {49}},\ \bibinfo {pages}
  {11560} (\bibinfo {year} {1994})}\BibitemShut {NoStop}%
\bibitem [{\citenamefont {Curtis}\ \emph {et~al.}(1954)\citenamefont {Curtis},
  \citenamefont {Doney},\ and\ \citenamefont {Johnson}}]{curtis54}%
  \BibitemOpen
  \bibfield  {author} {\bibinfo {author} {\bibfnamefont {C.~E.}\ \bibnamefont
  {Curtis}}, \bibinfo {author} {\bibfnamefont {L.~M.}\ \bibnamefont {Doney}}, \
  and\ \bibinfo {author} {\bibfnamefont {J.~R.}\ \bibnamefont {Johnson}},\
  }\href@noop {} {\bibfield  {journal} {\bibinfo  {journal} {J. Am. Ceram.
  Soc.}\ }\textbf {\bibinfo {volume} {37}},\ \bibinfo {pages} {458} (\bibinfo
  {year} {1954})}\BibitemShut {NoStop}%
\bibitem [{\citenamefont {Adam}\ and\ \citenamefont {Rodgers}(1959)}]{adam59}%
  \BibitemOpen
  \bibfield  {author} {\bibinfo {author} {\bibfnamefont {J.}~\bibnamefont
  {Adam}}\ and\ \bibinfo {author} {\bibfnamefont {M.~D.}\ \bibnamefont
  {Rodgers}},\ }\href@noop {} {\bibfield  {journal} {\bibinfo  {journal} {Acta.
  Crystallogr.}\ }\textbf {\bibinfo {volume} {12}},\ \bibinfo {pages} {951}
  (\bibinfo {year} {1959})}\BibitemShut {NoStop}%
\bibitem [{\citenamefont {Ruh}\ \emph {et~al.}(1968)\citenamefont {Ruh},
  \citenamefont {Garret}, \citenamefont {Domagala},\ and\ \citenamefont
  {Tallan}}]{ruh68}%
  \BibitemOpen
  \bibfield  {author} {\bibinfo {author} {\bibfnamefont {R.}~\bibnamefont
  {Ruh}}, \bibinfo {author} {\bibfnamefont {H.~J.}\ \bibnamefont {Garret}},
  \bibinfo {author} {\bibfnamefont {R.~F.}\ \bibnamefont {Domagala}}, \ and\
  \bibinfo {author} {\bibfnamefont {N.~M.}\ \bibnamefont {Tallan}},\
  }\href@noop {} {\bibfield  {journal} {\bibinfo  {journal} {J. Am. Ceram.
  Soc.}\ }\textbf {\bibinfo {volume} {51}},\ \bibinfo {pages} {23} (\bibinfo
  {year} {1968})}\BibitemShut {NoStop}%
\bibitem [{\citenamefont {Stacy}\ \emph {et~al.}(1972)\citenamefont {Stacy},
  \citenamefont {Johnstone},\ and\ \citenamefont {Wilder}}]{stacy72}%
  \BibitemOpen
  \bibfield  {author} {\bibinfo {author} {\bibfnamefont {D.~W.}\ \bibnamefont
  {Stacy}}, \bibinfo {author} {\bibfnamefont {J.~K.}\ \bibnamefont
  {Johnstone}}, \ and\ \bibinfo {author} {\bibfnamefont {D.~R.}\ \bibnamefont
  {Wilder}},\ }\href@noop {} {\bibfield  {journal} {\bibinfo  {journal} {J. Am.
  Ceram. Soc.}\ }\textbf {\bibinfo {volume} {51}},\ \bibinfo {pages} {482}
  (\bibinfo {year} {1972})}\BibitemShut {NoStop}%
\bibitem [{\citenamefont {Ruh}\ and\ \citenamefont {Patel}(1973)}]{ruh73}%
  \BibitemOpen
  \bibfield  {author} {\bibinfo {author} {\bibfnamefont {R.}~\bibnamefont
  {Ruh}}\ and\ \bibinfo {author} {\bibfnamefont {V.~A.}\ \bibnamefont
  {Patel}},\ }\href@noop {} {\bibfield  {journal} {\bibinfo  {journal} {J. Am.
  Ceram. Soc.}\ }\textbf {\bibinfo {volume} {56}},\ \bibinfo {pages} {606}
  (\bibinfo {year} {1973})}\BibitemShut {NoStop}%
\bibitem [{\citenamefont {Hann}\ \emph {et~al.}(1985)\citenamefont {Hann},
  \citenamefont {Suitch},\ and\ \citenamefont {Pentecost}}]{hann85}%
  \BibitemOpen
  \bibfield  {author} {\bibinfo {author} {\bibfnamefont {R.~E.}\ \bibnamefont
  {Hann}}, \bibinfo {author} {\bibfnamefont {P.~R.}\ \bibnamefont {Suitch}}, \
  and\ \bibinfo {author} {\bibfnamefont {J.~L.}\ \bibnamefont {Pentecost}},\
  }\href@noop {} {\bibfield  {journal} {\bibinfo  {journal} {J. Am. Ceram.
  Soc.}\ }\textbf {\bibinfo {volume} {68}},\ \bibinfo {pages} {C} (\bibinfo
  {year} {1985})}\BibitemShut {NoStop}%
\bibitem [{\citenamefont {Cisneros-Morales}\ and\ \citenamefont
  {Aita}(2010)}]{morales10}%
  \BibitemOpen
  \bibfield  {author} {\bibinfo {author} {\bibfnamefont {M.~C.}\ \bibnamefont
  {Cisneros-Morales}}\ and\ \bibinfo {author} {\bibfnamefont {C.~R.}\
  \bibnamefont {Aita}},\ }\href@noop {} {\bibfield  {journal} {\bibinfo
  {journal} {Appl. Phys. Lett.}\ }\textbf {\bibinfo {volume} {96}},\ \bibinfo
  {pages} {191904} (\bibinfo {year} {2010})}\BibitemShut {NoStop}%
\bibitem [{\citenamefont {El-Shanshoury}\ \emph {et~al.}(1970)\citenamefont
  {El-Shanshoury}, \citenamefont {Rudenko},\ and\ \citenamefont
  {Ibrahim}}]{shanshoury70}%
  \BibitemOpen
  \bibfield  {author} {\bibinfo {author} {\bibfnamefont {I.~A.}\ \bibnamefont
  {El-Shanshoury}}, \bibinfo {author} {\bibfnamefont {V.~A.}\ \bibnamefont
  {Rudenko}}, \ and\ \bibinfo {author} {\bibfnamefont {I.~A.}\ \bibnamefont
  {Ibrahim}},\ }\href@noop {} {\bibfield  {journal} {\bibinfo  {journal} {J.
  Am. Ceram. Soc.}\ }\textbf {\bibinfo {volume} {53}},\ \bibinfo {pages} {264}
  (\bibinfo {year} {1970})}\BibitemShut {NoStop}%
\bibitem [{\citenamefont {Wang}\ \emph {et~al.}(1992)\citenamefont {Wang},
  \citenamefont {Li},\ and\ \citenamefont {Stevens}}]{wang92}%
  \BibitemOpen
  \bibfield  {author} {\bibinfo {author} {\bibfnamefont {J.}~\bibnamefont
  {Wang}}, \bibinfo {author} {\bibfnamefont {H.~P.}\ \bibnamefont {Li}}, \ and\
  \bibinfo {author} {\bibfnamefont {R.}~\bibnamefont {Stevens}},\ }\href@noop
  {} {\bibfield  {journal} {\bibinfo  {journal} {J. Mater. Sci.}\ }\textbf
  {\bibinfo {volume} {27}},\ \bibinfo {pages} {5397} (\bibinfo {year}
  {1992})}\BibitemShut {NoStop}%
\bibitem [{\citenamefont {Dewhurst}\ and\ \citenamefont
  {Lowther}(1998)}]{dewhurst1998}%
  \BibitemOpen
  \bibfield  {author} {\bibinfo {author} {\bibfnamefont {J.}~\bibnamefont
  {Dewhurst}}\ and\ \bibinfo {author} {\bibfnamefont {J.}~\bibnamefont
  {Lowther}},\ }\href@noop {} {\bibfield  {journal} {\bibinfo  {journal}
  {Physical Review B}\ }\textbf {\bibinfo {volume} {57}},\ \bibinfo {pages}
  {741} (\bibinfo {year} {1998})}\BibitemShut {NoStop}%
\bibitem [{\citenamefont {Gallino}\ \emph {et~al.}(2011)\citenamefont
  {Gallino}, \citenamefont {Di~Valentin},\ and\ \citenamefont
  {Pacchioni}}]{gallino2011}%
  \BibitemOpen
  \bibfield  {author} {\bibinfo {author} {\bibfnamefont {F.}~\bibnamefont
  {Gallino}}, \bibinfo {author} {\bibfnamefont {C.}~\bibnamefont
  {Di~Valentin}}, \ and\ \bibinfo {author} {\bibfnamefont {G.}~\bibnamefont
  {Pacchioni}},\ }\href@noop {} {\bibfield  {journal} {\bibinfo  {journal}
  {Physical Chemistry Chemical Physics}\ }\textbf {\bibinfo {volume} {13}},\
  \bibinfo {pages} {17667} (\bibinfo {year} {2011})}\BibitemShut {NoStop}%
\bibitem [{\citenamefont {Jiang}\ \emph {et~al.}(2010)\citenamefont {Jiang},
  \citenamefont {Gomez-Abal}, \citenamefont {Rinke},\ and\ \citenamefont
  {Scheffler}}]{jiang2010}%
  \BibitemOpen
  \bibfield  {author} {\bibinfo {author} {\bibfnamefont {H.}~\bibnamefont
  {Jiang}}, \bibinfo {author} {\bibfnamefont {R.~I.}\ \bibnamefont
  {Gomez-Abal}}, \bibinfo {author} {\bibfnamefont {P.}~\bibnamefont {Rinke}}, \
  and\ \bibinfo {author} {\bibfnamefont {M.}~\bibnamefont {Scheffler}},\
  }\href@noop {} {\bibfield  {journal} {\bibinfo  {journal} {Physical Review
  B}\ }\textbf {\bibinfo {volume} {81}},\ \bibinfo {pages} {085119} (\bibinfo
  {year} {2010})}\BibitemShut {NoStop}%
\bibitem [{\citenamefont {Dash}\ \emph {et~al.}(2004)\citenamefont {Dash},
  \citenamefont {Vast}, \citenamefont {Baranek}, \citenamefont {Cheynet},\ and\
  \citenamefont {Reining}}]{dash2004}%
  \BibitemOpen
  \bibfield  {author} {\bibinfo {author} {\bibfnamefont {L.}~\bibnamefont
  {Dash}}, \bibinfo {author} {\bibfnamefont {N.}~\bibnamefont {Vast}}, \bibinfo
  {author} {\bibfnamefont {P.}~\bibnamefont {Baranek}}, \bibinfo {author}
  {\bibfnamefont {M.-C.}\ \bibnamefont {Cheynet}}, \ and\ \bibinfo {author}
  {\bibfnamefont {L.}~\bibnamefont {Reining}},\ }\href@noop {} {\bibfield
  {journal} {\bibinfo  {journal} {Physical Review B}\ }\textbf {\bibinfo
  {volume} {70}},\ \bibinfo {pages} {245116} (\bibinfo {year}
  {2004})}\BibitemShut {NoStop}%
\bibitem [{\citenamefont {Milman}\ \emph {et~al.}(2009)\citenamefont {Milman},
  \citenamefont {Perlov}, \citenamefont {Refson}, \citenamefont {Clark},
  \citenamefont {Gavartin},\ and\ \citenamefont {Winkler}}]{milman2009}%
  \BibitemOpen
  \bibfield  {author} {\bibinfo {author} {\bibfnamefont {V.}~\bibnamefont
  {Milman}}, \bibinfo {author} {\bibfnamefont {A.}~\bibnamefont {Perlov}},
  \bibinfo {author} {\bibfnamefont {K.}~\bibnamefont {Refson}}, \bibinfo
  {author} {\bibfnamefont {S.~J.}\ \bibnamefont {Clark}}, \bibinfo {author}
  {\bibfnamefont {J.}~\bibnamefont {Gavartin}}, \ and\ \bibinfo {author}
  {\bibfnamefont {B.}~\bibnamefont {Winkler}},\ }\href@noop {} {\bibfield
  {journal} {\bibinfo  {journal} {Journal of Physics: Condensed Matter}\
  }\textbf {\bibinfo {volume} {21}},\ \bibinfo {pages} {485404} (\bibinfo
  {year} {2009})}\BibitemShut {NoStop}%
\bibitem [{\citenamefont {Jansen}(1991)}]{jansen1991}%
  \BibitemOpen
  \bibfield  {author} {\bibinfo {author} {\bibfnamefont {H.~J.~F.}\
  \bibnamefont {Jansen}},\ }\href {\doibase 10.1103/PhysRevB.43.7267}
  {\bibfield  {journal} {\bibinfo  {journal} {Phys. Rev. B}\ }\textbf {\bibinfo
  {volume} {43}},\ \bibinfo {pages} {7267} (\bibinfo {year}
  {1991})}\BibitemShut {NoStop}%
\bibitem [{\citenamefont {Kr\'alik}\ \emph {et~al.}(1998)\citenamefont
  {Kr\'alik}, \citenamefont {Chang},\ and\ \citenamefont {Louie}}]{kralik1998}%
  \BibitemOpen
  \bibfield  {author} {\bibinfo {author} {\bibfnamefont {B.}~\bibnamefont
  {Kr\'alik}}, \bibinfo {author} {\bibfnamefont {E.~K.}\ \bibnamefont {Chang}},
  \ and\ \bibinfo {author} {\bibfnamefont {S.~G.}\ \bibnamefont {Louie}},\
  }\href {\doibase 10.1103/PhysRevB.57.7027} {\bibfield  {journal} {\bibinfo
  {journal} {Phys. Rev. B}\ }\textbf {\bibinfo {volume} {57}},\ \bibinfo
  {pages} {7027} (\bibinfo {year} {1998})}\BibitemShut {NoStop}%
\bibitem [{\citenamefont {Finnis}\ \emph {et~al.}(1998)\citenamefont {Finnis},
  \citenamefont {Paxton}, \citenamefont {Methfessel},\ and\ \citenamefont {van
  Schilfgaarde}}]{finnis1998}%
  \BibitemOpen
  \bibfield  {author} {\bibinfo {author} {\bibfnamefont {M.~W.}\ \bibnamefont
  {Finnis}}, \bibinfo {author} {\bibfnamefont {A.~T.}\ \bibnamefont {Paxton}},
  \bibinfo {author} {\bibfnamefont {M.}~\bibnamefont {Methfessel}}, \ and\
  \bibinfo {author} {\bibfnamefont {M.}~\bibnamefont {van Schilfgaarde}},\
  }\href {\doibase 10.1103/PhysRevLett.81.5149} {\bibfield  {journal} {\bibinfo
   {journal} {Phys. Rev. Lett.}\ }\textbf {\bibinfo {volume} {81}},\ \bibinfo
  {pages} {5149} (\bibinfo {year} {1998})}\BibitemShut {NoStop}%
\bibitem [{\citenamefont {Terki}\ \emph {et~al.}(2006)\citenamefont {Terki},
  \citenamefont {Bertrand}, \citenamefont {Aourag},\ and\ \citenamefont
  {Coddet}}]{terki2006}%
  \BibitemOpen
  \bibfield  {author} {\bibinfo {author} {\bibfnamefont {R.}~\bibnamefont
  {Terki}}, \bibinfo {author} {\bibfnamefont {G.}~\bibnamefont {Bertrand}},
  \bibinfo {author} {\bibfnamefont {H.}~\bibnamefont {Aourag}}, \ and\ \bibinfo
  {author} {\bibfnamefont {C.}~\bibnamefont {Coddet}},\ }\href@noop {}
  {\bibfield  {journal} {\bibinfo  {journal} {Materials science in
  semiconductor processing}\ }\textbf {\bibinfo {volume} {9}},\ \bibinfo
  {pages} {1006} (\bibinfo {year} {2006})}\BibitemShut {NoStop}%
\bibitem [{\citenamefont {Bredow}(2007)}]{bredow2007}%
  \BibitemOpen
  \bibfield  {author} {\bibinfo {author} {\bibfnamefont {T.}~\bibnamefont
  {Bredow}},\ }\href@noop {} {\bibfield  {journal} {\bibinfo  {journal}
  {Physical Review B}\ }\textbf {\bibinfo {volume} {75}},\ \bibinfo {pages}
  {144102} (\bibinfo {year} {2007})}\BibitemShut {NoStop}%
\bibitem [{\citenamefont {Daramola}\ \emph {et~al.}(2010)\citenamefont
  {Daramola}, \citenamefont {Muthuvel},\ and\ \citenamefont
  {Botte}}]{daramola2010}%
  \BibitemOpen
  \bibfield  {author} {\bibinfo {author} {\bibfnamefont {D.~A.}\ \bibnamefont
  {Daramola}}, \bibinfo {author} {\bibfnamefont {M.}~\bibnamefont {Muthuvel}},
  \ and\ \bibinfo {author} {\bibfnamefont {G.~G.}\ \bibnamefont {Botte}},\
  }\href@noop {} {\bibfield  {journal} {\bibinfo  {journal} {The Journal of
  Physical Chemistry B}\ }\textbf {\bibinfo {volume} {114}},\ \bibinfo {pages}
  {9323} (\bibinfo {year} {2010})}\BibitemShut {NoStop}%
\bibitem [{\citenamefont {Ricca}\ \emph {et~al.}(2015)\citenamefont {Ricca},
  \citenamefont {Ringued{\'e}}, \citenamefont {Cassir}, \citenamefont {Adamo},\
  and\ \citenamefont {Labat}}]{ricca2015}%
  \BibitemOpen
  \bibfield  {author} {\bibinfo {author} {\bibfnamefont {C.}~\bibnamefont
  {Ricca}}, \bibinfo {author} {\bibfnamefont {A.}~\bibnamefont {Ringued{\'e}}},
  \bibinfo {author} {\bibfnamefont {M.}~\bibnamefont {Cassir}}, \bibinfo
  {author} {\bibfnamefont {C.}~\bibnamefont {Adamo}}, \ and\ \bibinfo {author}
  {\bibfnamefont {F.}~\bibnamefont {Labat}},\ }\href@noop {} {\bibfield
  {journal} {\bibinfo  {journal} {Journal of computational chemistry}\ }\textbf
  {\bibinfo {volume} {36}},\ \bibinfo {pages} {9} (\bibinfo {year}
  {2015})}\BibitemShut {NoStop}%
\bibitem [{\citenamefont {Carbogno}\ \emph {et~al.}(2014)\citenamefont
  {Carbogno}, \citenamefont {Levi}, \citenamefont {Van~de Walle},\ and\
  \citenamefont {Scheffler}}]{carbogno2014}%
  \BibitemOpen
  \bibfield  {author} {\bibinfo {author} {\bibfnamefont {C.}~\bibnamefont
  {Carbogno}}, \bibinfo {author} {\bibfnamefont {C.~G.}\ \bibnamefont {Levi}},
  \bibinfo {author} {\bibfnamefont {C.~G.}\ \bibnamefont {Van~de Walle}}, \
  and\ \bibinfo {author} {\bibfnamefont {M.}~\bibnamefont {Scheffler}},\
  }\href@noop {} {\bibfield  {journal} {\bibinfo  {journal} {Physical Review
  B}\ }\textbf {\bibinfo {volume} {90}},\ \bibinfo {pages} {144109} (\bibinfo
  {year} {2014})}\BibitemShut {NoStop}%
\bibitem [{\citenamefont {Lowther}\ \emph {et~al.}(1999)\citenamefont
  {Lowther}, \citenamefont {Dewhurst}, \citenamefont {Leger},\ and\
  \citenamefont {Haines}}]{lowther99}%
  \BibitemOpen
  \bibfield  {author} {\bibinfo {author} {\bibfnamefont {J.~E.}\ \bibnamefont
  {Lowther}}, \bibinfo {author} {\bibfnamefont {J.~K.}\ \bibnamefont
  {Dewhurst}}, \bibinfo {author} {\bibfnamefont {J.~M.}\ \bibnamefont {Leger}},
  \ and\ \bibinfo {author} {\bibfnamefont {J.}~\bibnamefont {Haines}},\
  }\href@noop {} {\bibfield  {journal} {\bibinfo  {journal} {Phys. Rev. B}\
  }\textbf {\bibinfo {volume} {60}},\ \bibinfo {pages} {14485} (\bibinfo {year}
  {1999})}\BibitemShut {NoStop}%
\bibitem [{\citenamefont {Kang}\ \emph {et~al.}(2003)\citenamefont {Kang},
  \citenamefont {Lee},\ and\ \citenamefont {Chang}}]{kang03}%
  \BibitemOpen
  \bibfield  {author} {\bibinfo {author} {\bibfnamefont {J.}~\bibnamefont
  {Kang}}, \bibinfo {author} {\bibfnamefont {E.-C.}\ \bibnamefont {Lee}}, \
  and\ \bibinfo {author} {\bibfnamefont {K.~J.}\ \bibnamefont {Chang}},\
  }\href@noop {} {\bibfield  {journal} {\bibinfo  {journal} {Phys. Rev. B}\
  }\textbf {\bibinfo {volume} {68}},\ \bibinfo {pages} {054106} (\bibinfo
  {year} {2003})}\BibitemShut {NoStop}%
\bibitem [{\citenamefont {Jaffe}\ \emph {et~al.}(2005)\citenamefont {Jaffe},
  \citenamefont {Bachorz},\ and\ \citenamefont {Gutowski}}]{jaffe05}%
  \BibitemOpen
  \bibfield  {author} {\bibinfo {author} {\bibfnamefont {J.~E.}\ \bibnamefont
  {Jaffe}}, \bibinfo {author} {\bibfnamefont {R.~A.}\ \bibnamefont {Bachorz}},
  \ and\ \bibinfo {author} {\bibfnamefont {M.}~\bibnamefont {Gutowski}},\
  }\href@noop {} {\bibfield  {journal} {\bibinfo  {journal} {Phys. Rev. B}\
  }\textbf {\bibinfo {volume} {72}},\ \bibinfo {pages} {144107} (\bibinfo
  {year} {2005})}\BibitemShut {NoStop}%
\bibitem [{\citenamefont {Desgreniers}\ and\ \citenamefont
  {Lagarec}(1999)}]{desgreniers99}%
  \BibitemOpen
  \bibfield  {author} {\bibinfo {author} {\bibfnamefont {S.}~\bibnamefont
  {Desgreniers}}\ and\ \bibinfo {author} {\bibfnamefont {K.}~\bibnamefont
  {Lagarec}},\ }\href@noop {} {\bibfield  {journal} {\bibinfo  {journal} {Phys.
  Rev. B}\ }\textbf {\bibinfo {volume} {59}},\ \bibinfo {pages} {8467}
  (\bibinfo {year} {1999})}\BibitemShut {NoStop}%
\bibitem [{\citenamefont {Xiao}\ \emph {et~al.}(2014)\citenamefont {Xiao},
  \citenamefont {Sun}, \citenamefont {Ruzsinszky},\ and\ \citenamefont
  {Perdew}}]{perdew2014}%
  \BibitemOpen
  \bibfield  {author} {\bibinfo {author} {\bibfnamefont {B.}~\bibnamefont
  {Xiao}}, \bibinfo {author} {\bibfnamefont {J.}~\bibnamefont {Sun}}, \bibinfo
  {author} {\bibfnamefont {A.}~\bibnamefont {Ruzsinszky}}, \ and\ \bibinfo
  {author} {\bibfnamefont {J.~P.}\ \bibnamefont {Perdew}},\ }\href@noop {}
  {\bibfield  {journal} {\bibinfo  {journal} {Physical Review B}\ }\textbf
  {\bibinfo {volume} {90}},\ \bibinfo {pages} {085134} (\bibinfo {year}
  {2014})}\BibitemShut {NoStop}%
\bibitem [{\citenamefont {Hohenberg}\ and\ \citenamefont
  {Kohn}(1964)}]{Hohenberg1964}%
  \BibitemOpen
  \bibfield  {author} {\bibinfo {author} {\bibfnamefont {P.}~\bibnamefont
  {Hohenberg}}\ and\ \bibinfo {author} {\bibfnamefont {W.}~\bibnamefont
  {Kohn}},\ }\href {\doibase 10.1103/PhysRevB.7.1912} {\bibfield  {journal}
  {\bibinfo  {journal} {Phys. Rev.}\ }\textbf {\bibinfo {volume} {136}},\
  \bibinfo {pages} {B864} (\bibinfo {year} {1964})},\ \Eprint
  {http://arxiv.org/abs/1108.5632} {arXiv:1108.5632} \BibitemShut {NoStop}%
\bibitem [{\citenamefont {Kohn}\ and\ \citenamefont {Sham}(1965)}]{Kohn1965}%
  \BibitemOpen
  \bibfield  {author} {\bibinfo {author} {\bibfnamefont {W.}~\bibnamefont
  {Kohn}}\ and\ \bibinfo {author} {\bibfnamefont {L.~J.}\ \bibnamefont
  {Sham}},\ }\href {\doibase 10.1103/PhysRev.140.A1133} {\bibfield  {journal}
  {\bibinfo  {journal} {Physical Review}\ }\textbf {\bibinfo {volume} {140}},\
  \bibinfo {pages} {A1133} (\bibinfo {year} {1965})}\BibitemShut {NoStop}%
\bibitem [{\citenamefont {Lejaeghere}\ \emph {et~al.}(2016)\citenamefont
  {Lejaeghere}, \citenamefont {Bihlmayer}, \citenamefont {Bj{\"o}rkman},
  \citenamefont {Blaha}, \citenamefont {Bl{\"u}gel}, \citenamefont {Blum},
  \citenamefont {Caliste}, \citenamefont {Castelli}, \citenamefont {Clark},
  \citenamefont {Dal~Corso} \emph {et~al.}}]{DFT_Science2016}%
  \BibitemOpen
  \bibfield  {author} {\bibinfo {author} {\bibfnamefont {K.}~\bibnamefont
  {Lejaeghere}}, \bibinfo {author} {\bibfnamefont {G.}~\bibnamefont
  {Bihlmayer}}, \bibinfo {author} {\bibfnamefont {T.}~\bibnamefont
  {Bj{\"o}rkman}}, \bibinfo {author} {\bibfnamefont {P.}~\bibnamefont {Blaha}},
  \bibinfo {author} {\bibfnamefont {S.}~\bibnamefont {Bl{\"u}gel}}, \bibinfo
  {author} {\bibfnamefont {V.}~\bibnamefont {Blum}}, \bibinfo {author}
  {\bibfnamefont {D.}~\bibnamefont {Caliste}}, \bibinfo {author} {\bibfnamefont
  {I.~E.}\ \bibnamefont {Castelli}}, \bibinfo {author} {\bibfnamefont {S.~J.}\
  \bibnamefont {Clark}}, \bibinfo {author} {\bibfnamefont {A.}~\bibnamefont
  {Dal~Corso}},  \emph {et~al.},\ }\href@noop {} {\bibfield  {journal}
  {\bibinfo  {journal} {Science}\ }\textbf {\bibinfo {volume} {351}},\ \bibinfo
  {pages} {aad3000} (\bibinfo {year} {2016})}\BibitemShut {NoStop}%
\bibitem [{\citenamefont {Foulkes}\ \emph {et~al.}(2001)\citenamefont
  {Foulkes}, \citenamefont {Mitas}, \citenamefont {Needs},\ and\ \citenamefont
  {Rajagopal}}]{rev:qmcsolids}%
  \BibitemOpen
  \bibfield  {author} {\bibinfo {author} {\bibfnamefont {W.~M.~C.}\
  \bibnamefont {Foulkes}}, \bibinfo {author} {\bibfnamefont {L.}~\bibnamefont
  {Mitas}}, \bibinfo {author} {\bibfnamefont {R.~J.}\ \bibnamefont {Needs}}, \
  and\ \bibinfo {author} {\bibfnamefont {G.}~\bibnamefont {Rajagopal}},\ }\href
  {\doibase 10.1103/RevModPhys.73.33} {\bibfield  {journal} {\bibinfo
  {journal} {Reviews of Modern Physics}\ }\textbf {\bibinfo {volume} {73}},\
  \bibinfo {pages} {33} (\bibinfo {year} {2001})}\BibitemShut {NoStop}%
\bibitem [{\citenamefont {Blum}\ \emph {et~al.}(2009)\citenamefont {Blum},
  \citenamefont {Gehrke}, \citenamefont {Hanke}, \citenamefont {Havu},
  \citenamefont {Havu}, \citenamefont {Ren}, \citenamefont {Reuter},\ and\
  \citenamefont {Scheffler}}]{blum2009}%
  \BibitemOpen
  \bibfield  {author} {\bibinfo {author} {\bibfnamefont {V.}~\bibnamefont
  {Blum}}, \bibinfo {author} {\bibfnamefont {R.}~\bibnamefont {Gehrke}},
  \bibinfo {author} {\bibfnamefont {F.}~\bibnamefont {Hanke}}, \bibinfo
  {author} {\bibfnamefont {P.}~\bibnamefont {Havu}}, \bibinfo {author}
  {\bibfnamefont {V.}~\bibnamefont {Havu}}, \bibinfo {author} {\bibfnamefont
  {X.}~\bibnamefont {Ren}}, \bibinfo {author} {\bibfnamefont {K.}~\bibnamefont
  {Reuter}}, \ and\ \bibinfo {author} {\bibfnamefont {M.}~\bibnamefont
  {Scheffler}},\ }\href@noop {} {\bibfield  {journal} {\bibinfo  {journal}
  {Computer Physics Communications}\ }\textbf {\bibinfo {volume} {180}},\
  \bibinfo {pages} {2175} (\bibinfo {year} {2009})}\BibitemShut {NoStop}%
\bibitem [{\citenamefont {Ren}\ \emph {et~al.}(2012)\citenamefont {Ren},
  \citenamefont {Rinke}, \citenamefont {Blum}, \citenamefont {Wieferink},
  \citenamefont {Tkatchenko}, \citenamefont {Sanfilippo}, \citenamefont
  {Reuter},\ and\ \citenamefont {Scheffler}}]{ren2012}%
  \BibitemOpen
  \bibfield  {author} {\bibinfo {author} {\bibfnamefont {X.}~\bibnamefont
  {Ren}}, \bibinfo {author} {\bibfnamefont {P.}~\bibnamefont {Rinke}}, \bibinfo
  {author} {\bibfnamefont {V.}~\bibnamefont {Blum}}, \bibinfo {author}
  {\bibfnamefont {J.}~\bibnamefont {Wieferink}}, \bibinfo {author}
  {\bibfnamefont {A.}~\bibnamefont {Tkatchenko}}, \bibinfo {author}
  {\bibfnamefont {A.}~\bibnamefont {Sanfilippo}}, \bibinfo {author}
  {\bibfnamefont {K.}~\bibnamefont {Reuter}}, \ and\ \bibinfo {author}
  {\bibfnamefont {M.}~\bibnamefont {Scheffler}},\ }\href@noop {} {\bibfield
  {journal} {\bibinfo  {journal} {New Journal of Physics}\ }\textbf {\bibinfo
  {volume} {14}},\ \bibinfo {pages} {053020} (\bibinfo {year}
  {2012})}\BibitemShut {NoStop}%
\bibitem [{\citenamefont {Marek}\ \emph {et~al.}(2014)\citenamefont {Marek},
  \citenamefont {Blum}, \citenamefont {Johanni}, \citenamefont {Havu},
  \citenamefont {Lang}, \citenamefont {Auckenthaler}, \citenamefont {Heinecke},
  \citenamefont {Bungartz},\ and\ \citenamefont {Lederer}}]{marek2014}%
  \BibitemOpen
  \bibfield  {author} {\bibinfo {author} {\bibfnamefont {A.}~\bibnamefont
  {Marek}}, \bibinfo {author} {\bibfnamefont {V.}~\bibnamefont {Blum}},
  \bibinfo {author} {\bibfnamefont {R.}~\bibnamefont {Johanni}}, \bibinfo
  {author} {\bibfnamefont {V.}~\bibnamefont {Havu}}, \bibinfo {author}
  {\bibfnamefont {B.}~\bibnamefont {Lang}}, \bibinfo {author} {\bibfnamefont
  {T.}~\bibnamefont {Auckenthaler}}, \bibinfo {author} {\bibfnamefont
  {A.}~\bibnamefont {Heinecke}}, \bibinfo {author} {\bibfnamefont {H.-J.}\
  \bibnamefont {Bungartz}}, \ and\ \bibinfo {author} {\bibfnamefont
  {H.}~\bibnamefont {Lederer}},\ }\href@noop {} {\bibfield  {journal} {\bibinfo
   {journal} {Journal of Physics: Condensed Matter}\ }\textbf {\bibinfo
  {volume} {26}},\ \bibinfo {pages} {213201} (\bibinfo {year}
  {2014})}\BibitemShut {NoStop}%
\bibitem [{\citenamefont {Adamo}\ and\ \citenamefont
  {Barone}(1999)}]{adamo1999jchemphys}%
  \BibitemOpen
  \bibfield  {author} {\bibinfo {author} {\bibfnamefont {C.}~\bibnamefont
  {Adamo}}\ and\ \bibinfo {author} {\bibfnamefont {V.}~\bibnamefont {Barone}},\
  }\href@noop {} {\bibfield  {journal} {\bibinfo  {journal} {The Journal of
  Chemical Physics}\ }\textbf {\bibinfo {volume} {110}},\ \bibinfo {pages}
  {6158} (\bibinfo {year} {1999})}\BibitemShut {NoStop}%
\bibitem [{\citenamefont {Kim}\ \emph {et~al.}(2012)\citenamefont {Kim},
  \citenamefont {Esler}, \citenamefont {McMinis}, \citenamefont {Morales},
  \citenamefont {Clark}, \citenamefont {Shulenburger},\ and\ \citenamefont
  {Ceperley}}]{Kim2012}%
  \BibitemOpen
  \bibfield  {author} {\bibinfo {author} {\bibfnamefont {J.}~\bibnamefont
  {Kim}}, \bibinfo {author} {\bibfnamefont {K.~P.}\ \bibnamefont {Esler}},
  \bibinfo {author} {\bibfnamefont {J.}~\bibnamefont {McMinis}}, \bibinfo
  {author} {\bibfnamefont {M.~A.}\ \bibnamefont {Morales}}, \bibinfo {author}
  {\bibfnamefont {B.~K.}\ \bibnamefont {Clark}}, \bibinfo {author}
  {\bibfnamefont {L.}~\bibnamefont {Shulenburger}}, \ and\ \bibinfo {author}
  {\bibfnamefont {D.~M.}\ \bibnamefont {Ceperley}},\ }\href {\doibase
  10.1088/1742-6596/402/1/012008} {\bibfield  {journal} {\bibinfo  {journal}
  {Journal of Physics: Conference Series}\ }\textbf {\bibinfo {volume} {402}},\
  \bibinfo {pages} {012008} (\bibinfo {year} {2012})}\BibitemShut {NoStop}%
\bibitem [{\citenamefont {Esler}\ \emph {et~al.}(2012)\citenamefont {Esler},
  \citenamefont {Kim}, \citenamefont {Ceperley},\ and\ \citenamefont
  {Shulenburger}}]{Esler2012}%
  \BibitemOpen
  \bibfield  {author} {\bibinfo {author} {\bibfnamefont {K.}~\bibnamefont
  {Esler}}, \bibinfo {author} {\bibfnamefont {J.}~\bibnamefont {Kim}}, \bibinfo
  {author} {\bibfnamefont {D.}~\bibnamefont {Ceperley}}, \ and\ \bibinfo
  {author} {\bibfnamefont {L.}~\bibnamefont {Shulenburger}},\ }\href {\doibase
  10.1109/MCSE.2010.122} {\bibfield  {journal} {\bibinfo  {journal} {Computing
  in Science \& Engineering}\ }\textbf {\bibinfo {volume} {14}},\ \bibinfo
  {pages} {40} (\bibinfo {year} {2012})}\BibitemShut {NoStop}%
\bibitem [{\citenamefont {{\relax OPIUM Package}}()}]{opium}%
  \BibitemOpen
  \bibfield  {author} {\bibinfo {author} {\bibnamefont {{\relax OPIUM
  Package}}},\ }\href@noop {} {\bibinfo  {journal}
  {http://opium.sourceforge.net}\ }\BibitemShut {NoStop}%
\bibitem [{\citenamefont {Lide}(2003)}]{lide03}%
  \BibitemOpen
\bibfield  {journal} {  }\bibfield  {author} {\bibinfo {author} {\bibfnamefont
  {D.~R.}\ \bibnamefont {Lide}},\ }\href@noop {} {\emph {\bibinfo {title} {CRC
  Handbook of Chemistry and Physics, 84th Ed.}}}\ (\bibinfo  {publisher} {CRC
  Press},\ \bibinfo {year} {2003})\BibitemShut {NoStop}%
\bibitem [{\citenamefont {Feigerle}\ \emph {et~al.}(1998)\citenamefont
  {Feigerle}, \citenamefont {Corderman}, \citenamefont {Bobashev},\ and\
  \citenamefont {Lineberger}}]{feigerle98}%
  \BibitemOpen
  \bibfield  {author} {\bibinfo {author} {\bibfnamefont {C.~S.}\ \bibnamefont
  {Feigerle}}, \bibinfo {author} {\bibfnamefont {R.~R.}\ \bibnamefont
  {Corderman}}, \bibinfo {author} {\bibfnamefont {S.~V.}\ \bibnamefont
  {Bobashev}}, \ and\ \bibinfo {author} {\bibfnamefont {W.~C.}\ \bibnamefont
  {Lineberger}},\ }\href@noop {} {\bibfield  {journal} {\bibinfo  {journal}
  {Journal of Chemical Physics}\ }\textbf {\bibinfo {volume} {74}},\ \bibinfo
  {pages} {1580} (\bibinfo {year} {1998})}\BibitemShut {NoStop}%
\bibitem [{\citenamefont {Wang}\ \emph {et~al.}(2011)\citenamefont {Wang},
  \citenamefont {Zhang}, \citenamefont {Liu}, \citenamefont {Li},\ and\
  \citenamefont {Zhang}}]{wang11}%
  \BibitemOpen
  \bibfield  {author} {\bibinfo {author} {\bibfnamefont {B.-T.}\ \bibnamefont
  {Wang}}, \bibinfo {author} {\bibfnamefont {P.}~\bibnamefont {Zhang}},
  \bibinfo {author} {\bibfnamefont {H.-Y.}\ \bibnamefont {Liu}}, \bibinfo
  {author} {\bibfnamefont {W.-D.}\ \bibnamefont {Li}}, \ and\ \bibinfo {author}
  {\bibfnamefont {P.}~\bibnamefont {Zhang}},\ }\href@noop {} {\bibfield
  {journal} {\bibinfo  {journal} {Journal of Applied Physics}\ }\textbf
  {\bibinfo {volume} {109}},\ \bibinfo {pages} {063514} (\bibinfo {year}
  {2011})}\BibitemShut {NoStop}%
\bibitem [{\citenamefont {Zhao}\ \emph {et~al.}(2005)\citenamefont {Zhao},
  \citenamefont {Zhang}, \citenamefont {Pantea}, \citenamefont {Qian},
  \citenamefont {Daemen}, \citenamefont {Rigg}, \citenamefont {Hixson},
  \citenamefont {Gray~III}, \citenamefont {Yang}, \citenamefont {Wang},
  \citenamefont {Wang},\ and\ \citenamefont {Uchida}}]{zhao05}%
  \BibitemOpen
  \bibfield  {author} {\bibinfo {author} {\bibfnamefont {Y.}~\bibnamefont
  {Zhao}}, \bibinfo {author} {\bibfnamefont {J.}~\bibnamefont {Zhang}},
  \bibinfo {author} {\bibfnamefont {C.}~\bibnamefont {Pantea}}, \bibinfo
  {author} {\bibfnamefont {J.}~\bibnamefont {Qian}}, \bibinfo {author}
  {\bibfnamefont {L.~L.}\ \bibnamefont {Daemen}}, \bibinfo {author}
  {\bibfnamefont {P.~A.}\ \bibnamefont {Rigg}}, \bibinfo {author}
  {\bibfnamefont {R.~S.}\ \bibnamefont {Hixson}}, \bibinfo {author}
  {\bibfnamefont {G.~T.}\ \bibnamefont {Gray~III}}, \bibinfo {author}
  {\bibfnamefont {Y.}~\bibnamefont {Yang}}, \bibinfo {author} {\bibfnamefont
  {L.}~\bibnamefont {Wang}}, \bibinfo {author} {\bibfnamefont {Y.}~\bibnamefont
  {Wang}}, \ and\ \bibinfo {author} {\bibfnamefont {T.}~\bibnamefont
  {Uchida}},\ }\href@noop {} {\bibfield  {journal} {\bibinfo  {journal}
  {Physical Review B}\ }\textbf {\bibinfo {volume} {71}},\ \bibinfo {pages}
  {184119} (\bibinfo {year} {2005})}\BibitemShut {NoStop}%
\bibitem [{\citenamefont {Kittel}(2005)}]{kittel05}%
  \BibitemOpen
  \bibfield  {author} {\bibinfo {author} {\bibfnamefont {C.}~\bibnamefont
  {Kittel}},\ }\href@noop {} {\emph {\bibinfo {title} {Introduction to Solid
  State Physics, 8th Ed.}}}\ (\bibinfo  {publisher} {John Wiley \& Sons,
  Inc.},\ \bibinfo {year} {2005})\BibitemShut {NoStop}%
\bibitem [{\citenamefont {Russell}(1953)}]{russell53}%
  \BibitemOpen
  \bibfield  {author} {\bibinfo {author} {\bibfnamefont {R.~B.}\ \bibnamefont
  {Russell}},\ }\href@noop {} {\bibfield  {journal} {\bibinfo  {journal} {J.\
  Appl.\ Phys.}\ }\textbf {\bibinfo {volume} {24}},\ \bibinfo {pages} {232}
  (\bibinfo {year} {1953})}\BibitemShut {NoStop}%
\bibitem [{\citenamefont {Lin}\ \emph {et~al.}(2001)\citenamefont {Lin},
  \citenamefont {Zong},\ and\ \citenamefont {Ceperley}}]{Ceperley:twists}%
  \BibitemOpen
  \bibfield  {author} {\bibinfo {author} {\bibfnamefont {C.}~\bibnamefont
  {Lin}}, \bibinfo {author} {\bibfnamefont {F.~H.}\ \bibnamefont {Zong}}, \
  and\ \bibinfo {author} {\bibfnamefont {D.~M.}\ \bibnamefont {Ceperley}},\
  }\href {\doibase 10.1103/PhysRevE.64.016702} {\bibfield  {journal} {\bibinfo
  {journal} {Physical review. E, Statistical, nonlinear, and soft matter
  physics}\ }\textbf {\bibinfo {volume} {64}},\ \bibinfo {pages} {016702}
  (\bibinfo {year} {2001})}\BibitemShut {NoStop}%
\bibitem [{\citenamefont {Vinet}\ \emph {et~al.}(1986)\citenamefont {Vinet},
  \citenamefont {Farrante}, \citenamefont {Smith},\ and\ \citenamefont
  {Rose}}]{vinet86}%
  \BibitemOpen
  \bibfield  {author} {\bibinfo {author} {\bibfnamefont {P.}~\bibnamefont
  {Vinet}}, \bibinfo {author} {\bibfnamefont {J.}~\bibnamefont {Farrante}},
  \bibinfo {author} {\bibfnamefont {J.~R.}\ \bibnamefont {Smith}}, \ and\
  \bibinfo {author} {\bibfnamefont {J.~H.}\ \bibnamefont {Rose}},\ }\href@noop
  {} {\bibfield  {journal} {\bibinfo  {journal} {Journal of Physics C : Solid
  State Physics}\ }\textbf {\bibinfo {volume} {19}},\ \bibinfo {pages} {L467}
  (\bibinfo {year} {1986})}\BibitemShut {NoStop}%
\bibitem [{\citenamefont {Benali}\ \emph {et~al.}(2016)\citenamefont {Benali},
  \citenamefont {Shulenburger}, \citenamefont {Krogel}, \citenamefont {Zhong},
  \citenamefont {Kent},\ and\ \citenamefont {Heinonen}}]{BenaliTi4O7}%
  \BibitemOpen
  \bibfield  {author} {\bibinfo {author} {\bibfnamefont {A.}~\bibnamefont
  {Benali}}, \bibinfo {author} {\bibfnamefont {L.}~\bibnamefont
  {Shulenburger}}, \bibinfo {author} {\bibfnamefont {J.~T.}\ \bibnamefont
  {Krogel}}, \bibinfo {author} {\bibfnamefont {X.}~\bibnamefont {Zhong}},
  \bibinfo {author} {\bibfnamefont {P.~R.~C.}\ \bibnamefont {Kent}}, \ and\
  \bibinfo {author} {\bibfnamefont {O.}~\bibnamefont {Heinonen}},\ }\href@noop
  {} {\bibfield  {journal} {\bibinfo  {journal} {Phys.\ Chem.\ Chem.\ Phys.}\
  }\textbf {\bibinfo {volume} {18}},\ \bibinfo {pages} {18323} (\bibinfo {year}
  {2016})}\BibitemShut {NoStop}%
\bibitem [{\citenamefont {Luo}\ \emph {et~al.}(2016)\citenamefont {Luo},
  \citenamefont {Benali}, \citenamefont {Shulenburger}, \citenamefont {Krogel},
  \citenamefont {Heinonen},\ and\ \citenamefont {Kent}}]{shin17}%
  \BibitemOpen
  \bibfield  {author} {\bibinfo {author} {\bibfnamefont {Y.}~\bibnamefont
  {Luo}}, \bibinfo {author} {\bibfnamefont {A.}~\bibnamefont {Benali}},
  \bibinfo {author} {\bibfnamefont {L.}~\bibnamefont {Shulenburger}}, \bibinfo
  {author} {\bibfnamefont {J.~T.}\ \bibnamefont {Krogel}}, \bibinfo {author}
  {\bibfnamefont {O.}~\bibnamefont {Heinonen}}, \ and\ \bibinfo {author}
  {\bibfnamefont {P.~R.~C.}\ \bibnamefont {Kent}},\ }\href@noop {} {\bibfield
  {journal} {\bibinfo  {journal} {New\ J.\ Phys.}\ }\textbf {\bibinfo {volume}
  {18}},\ \bibinfo {pages} {113049} (\bibinfo {year} {2016})}\BibitemShut
  {NoStop}%
\bibitem [{\citenamefont {Shin}\ \emph {et~al.}(2017)\citenamefont {Shin},
  \citenamefont {Luo}, \citenamefont {Ganesh}, \citenamefont {Balachandran},
  \citenamefont {Krogel}, \citenamefont {Kent}, \citenamefont {Benali},\ and\
  \citenamefont {Heinonen}}]{shin17NiO}%
  \BibitemOpen
  \bibfield  {author} {\bibinfo {author} {\bibfnamefont {H.}~\bibnamefont
  {Shin}}, \bibinfo {author} {\bibfnamefont {Y.}~\bibnamefont {Luo}}, \bibinfo
  {author} {\bibfnamefont {P.}~\bibnamefont {Ganesh}}, \bibinfo {author}
  {\bibfnamefont {J.}~\bibnamefont {Balachandran}}, \bibinfo {author}
  {\bibfnamefont {J.~T.}\ \bibnamefont {Krogel}}, \bibinfo {author}
  {\bibfnamefont {P.~R.~C.}\ \bibnamefont {Kent}}, \bibinfo {author}
  {\bibfnamefont {A.}~\bibnamefont {Benali}}, \ and\ \bibinfo {author}
  {\bibfnamefont {O.}~\bibnamefont {Heinonen}},\ }\href@noop {} {\bibfield
  {journal} {\bibinfo  {journal} {Phys.\ Rev.\ Materials}\ }\textbf {\bibinfo
  {volume} {1}},\ \bibinfo {pages} {073603} (\bibinfo {year}
  {2017})}\BibitemShut {NoStop}%
\bibitem [{\citenamefont {Fabris}\ \emph {et~al.}(2001)\citenamefont {Fabris},
  \citenamefont {Paxton},\ and\ \citenamefont {Finnis}}]{fabris2001}%
  \BibitemOpen
  \bibfield  {author} {\bibinfo {author} {\bibfnamefont {S.}~\bibnamefont
  {Fabris}}, \bibinfo {author} {\bibfnamefont {A.~T.}\ \bibnamefont {Paxton}},
  \ and\ \bibinfo {author} {\bibfnamefont {M.~W.}\ \bibnamefont {Finnis}},\
  }\href {\doibase 10.1103/PhysRevB.63.094101} {\bibfield  {journal} {\bibinfo
  {journal} {Phys. Rev. B}\ }\textbf {\bibinfo {volume} {63}},\ \bibinfo
  {pages} {094101} (\bibinfo {year} {2001})}\BibitemShut {NoStop}%
\bibitem [{\citenamefont {Sternik}\ and\ \citenamefont
  {Parlinski}(2005)}]{sternik2005}%
  \BibitemOpen
  \bibfield  {author} {\bibinfo {author} {\bibfnamefont {M.}~\bibnamefont
  {Sternik}}\ and\ \bibinfo {author} {\bibfnamefont {K.}~\bibnamefont
  {Parlinski}},\ }\href@noop {} {\bibfield  {journal} {\bibinfo  {journal} {The
  Journal of chemical physics}\ }\textbf {\bibinfo {volume} {123}},\ \bibinfo
  {pages} {204708} (\bibinfo {year} {2005})}\BibitemShut {NoStop}%
\bibitem [{\citenamefont {Bouvier}\ \emph {et~al.}(2001)\citenamefont
  {Bouvier}, \citenamefont {Djurado}, \citenamefont {Ritter}, \citenamefont
  {Dianoux},\ and\ \citenamefont {Lucazeau}}]{bouvier01}%
  \BibitemOpen
  \bibfield  {author} {\bibinfo {author} {\bibfnamefont {P.}~\bibnamefont
  {Bouvier}}, \bibinfo {author} {\bibfnamefont {E.}~\bibnamefont {Djurado}},
  \bibinfo {author} {\bibfnamefont {C.}~\bibnamefont {Ritter}}, \bibinfo
  {author} {\bibfnamefont {A.~J.}\ \bibnamefont {Dianoux}}, \ and\ \bibinfo
  {author} {\bibfnamefont {G.}~\bibnamefont {Lucazeau}},\ }\href@noop {}
  {\bibfield  {journal} {\bibinfo  {journal} {Int. J. Inorg. Mater.}\ }\textbf
  {\bibinfo {volume} {3}},\ \bibinfo {pages} {647} (\bibinfo {year}
  {2001})}\BibitemShut {NoStop}%
\bibitem [{\citenamefont {Bouvier}\ \emph {et~al.}(2000)\citenamefont
  {Bouvier}, \citenamefont {Djurado}, \citenamefont {Lucazeau},\ and\
  \citenamefont {Bihan}}]{bouvier00}%
  \BibitemOpen
  \bibfield  {author} {\bibinfo {author} {\bibfnamefont {P.}~\bibnamefont
  {Bouvier}}, \bibinfo {author} {\bibfnamefont {E.}~\bibnamefont {Djurado}},
  \bibinfo {author} {\bibfnamefont {G.}~\bibnamefont {Lucazeau}}, \ and\
  \bibinfo {author} {\bibfnamefont {T.~L.}\ \bibnamefont {Bihan}},\ }\href@noop
  {} {\bibfield  {journal} {\bibinfo  {journal} {Phys. Rev. B}\ }\textbf
  {\bibinfo {volume} {62}},\ \bibinfo {pages} {8731} (\bibinfo {year}
  {2000})}\BibitemShut {NoStop}%
\bibitem [{\citenamefont {Kandil}\ \emph {et~al.}(1984)\citenamefont {Kandil},
  \citenamefont {Greiner},\ and\ \citenamefont {Smith}}]{kandil1984}%
  \BibitemOpen
  \bibfield  {author} {\bibinfo {author} {\bibfnamefont {H.}~\bibnamefont
  {Kandil}}, \bibinfo {author} {\bibfnamefont {J.}~\bibnamefont {Greiner}}, \
  and\ \bibinfo {author} {\bibfnamefont {J.}~\bibnamefont {Smith}},\
  }\href@noop {} {\bibfield  {journal} {\bibinfo  {journal} {Journal of the
  American Ceramic Society}\ }\textbf {\bibinfo {volume} {67}},\ \bibinfo
  {pages} {341} (\bibinfo {year} {1984})}\BibitemShut {NoStop}%
\bibitem [{\citenamefont {Fukuhara}\ and\ \citenamefont
  {Yamauchi}(1993)}]{fukuhara93}%
  \BibitemOpen
  \bibfield  {author} {\bibinfo {author} {\bibfnamefont {M.}~\bibnamefont
  {Fukuhara}}\ and\ \bibinfo {author} {\bibfnamefont {I.}~\bibnamefont
  {Yamauchi}},\ }\href@noop {} {\bibfield  {journal} {\bibinfo  {journal} {J.
  Mater. Sci.}\ }\textbf {\bibinfo {volume} {28}},\ \bibinfo {pages} {4681}
  (\bibinfo {year} {1993})}\BibitemShut {NoStop}%
\bibitem [{\citenamefont {Bouvier}\ \emph {et~al.}(2003)\citenamefont
  {Bouvier}, \citenamefont {Dmitriev},\ and\ \citenamefont
  {Lucazeau}}]{bouvier03}%
  \BibitemOpen
  \bibfield  {author} {\bibinfo {author} {\bibfnamefont {P.}~\bibnamefont
  {Bouvier}}, \bibinfo {author} {\bibfnamefont {V.}~\bibnamefont {Dmitriev}}, \
  and\ \bibinfo {author} {\bibfnamefont {G.}~\bibnamefont {Lucazeau}},\
  }\href@noop {} {\bibfield  {journal} {\bibinfo  {journal} {Eur. Phys. J. B}\
  }\textbf {\bibinfo {volume} {35}},\ \bibinfo {pages} {301} (\bibinfo {year}
  {2003})}\BibitemShut {NoStop}%
\bibitem [{\citenamefont {Lynch}(1974)}]{lynch74}%
  \BibitemOpen
  \bibfield  {author} {\bibinfo {author} {\bibfnamefont {C.~T.}\ \bibnamefont
  {Lynch}},\ }\href@noop {} {\emph {\bibinfo {title} {CRC Handbook of Materials
  Science}}}\ (\bibinfo  {publisher} {CRC Press},\ \bibinfo {year}
  {1974})\BibitemShut {NoStop}%
\bibitem [{\citenamefont {Adams}\ \emph {et~al.}(1991)\citenamefont {Adams},
  \citenamefont {Leonard}, \citenamefont {Russel},\ and\ \citenamefont
  {Cernik}}]{adams91}%
  \BibitemOpen
  \bibfield  {author} {\bibinfo {author} {\bibfnamefont {D.~M.}\ \bibnamefont
  {Adams}}, \bibinfo {author} {\bibfnamefont {S.}~\bibnamefont {Leonard}},
  \bibinfo {author} {\bibfnamefont {D.~R.}\ \bibnamefont {Russel}}, \ and\
  \bibinfo {author} {\bibfnamefont {R.~J.}\ \bibnamefont {Cernik}},\
  }\href@noop {} {\bibfield  {journal} {\bibinfo  {journal} {J. Phys. Chem.
  Solids}\ }\textbf {\bibinfo {volume} {52}},\ \bibinfo {pages} {1181}
  (\bibinfo {year} {1991})}\BibitemShut {NoStop}%
\bibitem [{\citenamefont {Ruh}\ and\ \citenamefont {Corfield}(1970)}]{ruh70}%
  \BibitemOpen
  \bibfield  {author} {\bibinfo {author} {\bibfnamefont {R.}~\bibnamefont
  {Ruh}}\ and\ \bibinfo {author} {\bibfnamefont {P.~W.~R.}\ \bibnamefont
  {Corfield}},\ }\href@noop {} {\bibfield  {journal} {\bibinfo  {journal} {J.
  Am. Ceram. Soc.}\ }\textbf {\bibinfo {volume} {53}},\ \bibinfo {pages} {126}
  (\bibinfo {year} {1970})}\BibitemShut {NoStop}%
\bibitem [{\citenamefont {Terki}\ \emph {et~al.}(2008)\citenamefont {Terki},
  \citenamefont {Bertrand}, \citenamefont {Aourag},\ and\ \citenamefont
  {Coddet}}]{terki08}%
  \BibitemOpen
  \bibfield  {author} {\bibinfo {author} {\bibfnamefont {R.}~\bibnamefont
  {Terki}}, \bibinfo {author} {\bibfnamefont {G.}~\bibnamefont {Bertrand}},
  \bibinfo {author} {\bibfnamefont {H.}~\bibnamefont {Aourag}}, \ and\ \bibinfo
  {author} {\bibfnamefont {C.}~\bibnamefont {Coddet}},\ }\href@noop {}
  {\bibfield  {journal} {\bibinfo  {journal} {Mater. Lett.}\ }\textbf {\bibinfo
  {volume} {62}},\ \bibinfo {pages} {1484} (\bibinfo {year}
  {2008})}\BibitemShut {NoStop}%
\bibitem [{\citenamefont {Foster}\ \emph {et~al.}(2002)\citenamefont {Foster},
  \citenamefont {Gejo}, \citenamefont {Shluger},\ and\ \citenamefont
  {Nieminen}}]{foster02}%
  \BibitemOpen
  \bibfield  {author} {\bibinfo {author} {\bibfnamefont {A.~S.}\ \bibnamefont
  {Foster}}, \bibinfo {author} {\bibfnamefont {F.~L.}\ \bibnamefont {Gejo}},
  \bibinfo {author} {\bibfnamefont {A.~L.}\ \bibnamefont {Shluger}}, \ and\
  \bibinfo {author} {\bibfnamefont {R.~M.}\ \bibnamefont {Nieminen}},\
  }\href@noop {} {\bibfield  {journal} {\bibinfo  {journal} {Phys. Rev. B}\
  }\textbf {\bibinfo {volume} {65}},\ \bibinfo {pages} {174117} (\bibinfo
  {year} {2002})}\BibitemShut {NoStop}%
\bibitem [{\citenamefont {Iskandarova}\ \emph {et~al.}(2003)\citenamefont
  {Iskandarova}, \citenamefont {Knizhnik}, \citenamefont {Rykova},
  \citenamefont {Bagatur'yants}, \citenamefont {Potapkin},\ and\ \citenamefont
  {Korkin}}]{Iskandarova03}%
  \BibitemOpen
  \bibfield  {author} {\bibinfo {author} {\bibfnamefont {I.~M.}\ \bibnamefont
  {Iskandarova}}, \bibinfo {author} {\bibfnamefont {A.~A.}\ \bibnamefont
  {Knizhnik}}, \bibinfo {author} {\bibfnamefont {E.~A.}\ \bibnamefont
  {Rykova}}, \bibinfo {author} {\bibfnamefont {A.~A.}\ \bibnamefont
  {Bagatur'yants}}, \bibinfo {author} {\bibfnamefont {B.~V.}\ \bibnamefont
  {Potapkin}}, \ and\ \bibinfo {author} {\bibfnamefont {A.~A.}\ \bibnamefont
  {Korkin}},\ }\href@noop {} {\bibfield  {journal} {\bibinfo  {journal}
  {Microelectron. Eng.}\ }\textbf {\bibinfo {volume} {69}},\ \bibinfo {pages}
  {587} (\bibinfo {year} {2003})}\BibitemShut {NoStop}%
\bibitem [{\citenamefont {Li}\ \emph {et~al.}(2013)\citenamefont {Li},
  \citenamefont {Han}, \citenamefont {Meng}, \citenamefont {Lu},\ and\
  \citenamefont {Tohyama}}]{li13}%
  \BibitemOpen
  \bibfield  {author} {\bibinfo {author} {\bibfnamefont {J.}~\bibnamefont
  {Li}}, \bibinfo {author} {\bibfnamefont {J.}~\bibnamefont {Han}}, \bibinfo
  {author} {\bibfnamefont {S.}~\bibnamefont {Meng}}, \bibinfo {author}
  {\bibfnamefont {H.}~\bibnamefont {Lu}}, \ and\ \bibinfo {author}
  {\bibfnamefont {T.}~\bibnamefont {Tohyama}},\ }\href@noop {} {\bibfield
  {journal} {\bibinfo  {journal} {Appl. Phys. Lett.}\ }\textbf {\bibinfo
  {volume} {103}},\ \bibinfo {pages} {071916} (\bibinfo {year}
  {2013})}\BibitemShut {NoStop}%
\bibitem [{\citenamefont {Beltr{\'a}n}\ \emph {et~al.}(2008)\citenamefont
  {Beltr{\'a}n}, \citenamefont {Mu{\~n}oz},\ and\ \citenamefont
  {Hafner}}]{beltran2008}%
  \BibitemOpen
  \bibfield  {author} {\bibinfo {author} {\bibfnamefont {J.}~\bibnamefont
  {Beltr{\'a}n}}, \bibinfo {author} {\bibfnamefont {M.}~\bibnamefont
  {Mu{\~n}oz}}, \ and\ \bibinfo {author} {\bibfnamefont {J.}~\bibnamefont
  {Hafner}},\ }\href@noop {} {\bibfield  {journal} {\bibinfo  {journal} {New
  Journal of Physics}\ }\textbf {\bibinfo {volume} {10}},\ \bibinfo {pages}
  {063031} (\bibinfo {year} {2008})}\BibitemShut {NoStop}%
\bibitem [{\citenamefont {Lyons}\ \emph {et~al.}(2011)\citenamefont {Lyons},
  \citenamefont {Janotti},\ and\ \citenamefont {Van~de Walle}}]{lyons2011}%
  \BibitemOpen
  \bibfield  {author} {\bibinfo {author} {\bibfnamefont {J.}~\bibnamefont
  {Lyons}}, \bibinfo {author} {\bibfnamefont {A.}~\bibnamefont {Janotti}}, \
  and\ \bibinfo {author} {\bibfnamefont {C.}~\bibnamefont {Van~de Walle}},\
  }\href@noop {} {\bibfield  {journal} {\bibinfo  {journal} {Microelectronic
  Engineering}\ }\textbf {\bibinfo {volume} {88}},\ \bibinfo {pages} {1452}
  (\bibinfo {year} {2011})}\BibitemShut {NoStop}%
\bibitem [{\citenamefont {Perry}(2011)}]{perry11}%
  \BibitemOpen
  \bibfield  {author} {\bibinfo {author} {\bibfnamefont {D.~L.}\ \bibnamefont
  {Perry}},\ }\href@noop {} {\emph {\bibinfo {title} {Handbook of Inorganic
  Compounds, 2nd Ed.}}}\ (\bibinfo  {publisher} {CRC Press},\ \bibinfo {year}
  {2011})\BibitemShut {NoStop}%
\bibitem [{\citenamefont {Haggerty}\ \emph {et~al.}(2014)\citenamefont
  {Haggerty}, \citenamefont {Sarin}, \citenamefont {Apostolov}, \citenamefont
  {Driemeyer},\ and\ \citenamefont {Kriven}}]{haggerty14}%
  \BibitemOpen
  \bibfield  {author} {\bibinfo {author} {\bibfnamefont {R.~P.}\ \bibnamefont
  {Haggerty}}, \bibinfo {author} {\bibfnamefont {P.}~\bibnamefont {Sarin}},
  \bibinfo {author} {\bibfnamefont {Z.~D.}\ \bibnamefont {Apostolov}}, \bibinfo
  {author} {\bibfnamefont {P.~E.}\ \bibnamefont {Driemeyer}}, \ and\ \bibinfo
  {author} {\bibfnamefont {W.~M.}\ \bibnamefont {Kriven}},\ }\href@noop {}
  {\bibfield  {journal} {\bibinfo  {journal} {J. Am. Ceram. Soc.}\ }\textbf
  {\bibinfo {volume} {97}},\ \bibinfo {pages} {2213} (\bibinfo {year}
  {2014})}\BibitemShut {NoStop}%
\bibitem [{\citenamefont {Glushkova}\ and\ \citenamefont
  {Kravchinskaya}(1985)}]{glushkova85}%
  \BibitemOpen
  \bibfield  {author} {\bibinfo {author} {\bibfnamefont {V.~B.}\ \bibnamefont
  {Glushkova}}\ and\ \bibinfo {author} {\bibfnamefont {M.~V.}\ \bibnamefont
  {Kravchinskaya}},\ }\href@noop {} {\bibfield  {journal} {\bibinfo  {journal}
  {Ceram. Intl.}\ }\textbf {\bibinfo {volume} {11}},\ \bibinfo {pages} {56}
  (\bibinfo {year} {1985})}\BibitemShut {NoStop}%
\bibitem [{\citenamefont {Senft}\ and\ \citenamefont
  {Stubican}(1983)}]{senft83}%
  \BibitemOpen
  \bibfield  {author} {\bibinfo {author} {\bibfnamefont {G.~B.}\ \bibnamefont
  {Senft}}\ and\ \bibinfo {author} {\bibfnamefont {V.~S.}\ \bibnamefont
  {Stubican}},\ }\href@noop {} {\bibfield  {journal} {\bibinfo  {journal} {Mat.
  Res. Bull.}\ }\textbf {\bibinfo {volume} {18}},\ \bibinfo {pages} {1163}
  (\bibinfo {year} {1983})}\BibitemShut {NoStop}%
\bibitem [{\citenamefont {Buckley}(1968)}]{buckley68}%
  \BibitemOpen
  \bibfield  {author} {\bibinfo {author} {\bibfnamefont {J.~D.}\ \bibnamefont
  {Buckley}},\ }\emph {\bibinfo {title} {Stabilization of the phase
  transformations in hafnium oxide}},\ \href@noop {} {Ph.D. thesis},\ \bibinfo
  {school} {Iowa State University}, \bibinfo {address} {Ames, IA, US} (\bibinfo
  {year} {1968})\BibitemShut {NoStop}%
\bibitem [{\citenamefont {Grau-Crespo}\ \emph {et~al.}(2012)\citenamefont
  {Grau-Crespo}, \citenamefont {Wang},\ and\ \citenamefont
  {Schwingenschl{\"o}gl}}]{grau2012}%
  \BibitemOpen
  \bibfield  {author} {\bibinfo {author} {\bibfnamefont {R.}~\bibnamefont
  {Grau-Crespo}}, \bibinfo {author} {\bibfnamefont {H.}~\bibnamefont {Wang}}, \
  and\ \bibinfo {author} {\bibfnamefont {U.}~\bibnamefont
  {Schwingenschl{\"o}gl}},\ }\href@noop {} {\bibfield  {journal} {\bibinfo
  {journal} {Physical Review B}\ }\textbf {\bibinfo {volume} {86}},\ \bibinfo
  {pages} {081101} (\bibinfo {year} {2012})}\BibitemShut {NoStop}%
\end{thebibliography}%

\end{document}